\documentclass[twocolumn,aps,prd,superscriptaddress,nofootinbib,floatfix,preprintnumbers]{revtex4-1}

\usepackage{graphicx,multirow}
\usepackage{xspace}

\usepackage{amsmath,amsfonts}
\allowdisplaybreaks

\usepackage{hyperref}

\newcommand{\nn}{\nonumber}
\newcommand{\beq}{\begin{equation}}
\newcommand{\eeq}{\end{equation}}
\newcommand{\beqa}{\begin{eqnarray}}
\newcommand{\eeqa}{\end{eqnarray}}

\newcommand{\GeV}{\text{GeV}}
\newcommand{\MeV}{\text{MeV}}

\newcommand{\babar}{\mbox{\ensuremath{{\displaystyle B}\!{\scriptstyle A}{\displaystyle B}\!{\scriptstyle AR}}}\xspace}

\def\d{{\rm d}}
\newcommand{\ov}{\overline}
\newcommand{\ds}{\displaystyle}

\newcommand{\lqcd}{\ensuremath{\Lambda_{\rm QCD}}\xspace}

\newcommand{\Bbar}{\,\overline{\!B}{}}
\newcommand{\Dbar}{\,\overline{\!D}{}}
\newcommand{\Kbar}{\,\overline{\!K}{}}
\def\B0bar{\Bbar{}^0}
\def\D0bar{\Dbar{}^0}
\def\K0bar{\Kbar{}^0}

\def\rt{\rho_\tau}
\def\rl{\rho_\ell}
\def\ApproxA{Approximation~A\xspace}
\def\ApproxB{Approximation~B\xspace}
\def\ApproxC{Approximation~C\xspace}

\makeatletter
\g@addto@macro\bfseries{\boldmath}
\makeatother

\begin{document}


\title{Semileptonic $B_{(s)}$ decays to excited charmed mesons with $e,\mu,\tau$
\\
and searching for new physics with $R(D^{**})$}

\author{Florian U.\ Bernlochner}
\affiliation{Physikalisches Institut der Rheinischen Friedrich-Wilhelms-Universit\"at Bonn, 53115 Bonn, Germany}

\author{Zoltan Ligeti}
\affiliation{Ernest Orlando Lawrence Berkeley National Laboratory,
University of California, Berkeley, CA 94720}

\begin{abstract}

Semileptonic $B$ meson decays into the four lightest excited charmed meson
states ($D_0^*$, $D_1^*$, $D_1$, and $D_2^*$) and their counterparts with $s$
quarks are investigated, including the full lepton mass dependence.  We derive
the standard model predictions for the differential branching fractions, as well
as predictions for the ratios of the semi-tauonic and light lepton semileptonic
branching fractions.  These can be systematically improved using future
measurements of the total or differential semileptonic rates to $e$ and $\mu$,
as well as the two-body hadronic branching fractions with a pion, related by
factorization to the semileptonic rate at maximal recoil.  To illustrate the
different sensitivities to new physics, we explore the dependence of the ratio
of semi-tauonic and light-lepton branching fractions on the type-II and type-III
two-Higgs-doublet model parameters, $\tan\beta$ and $m_{H}^\pm$, for all four
states. 

\end{abstract}

\maketitle

\section{Introduction}

The study of semileptonic $b \to c$ decays has been a central focus of the $B$
factory experiments \babar and Belle, as well as LHCb.  Such decays are
important for the measurement of the Cabibbo-Kobayashi-Maskawa (CKM) matrix
element $|V_{cb}|$ and are also probes of physics beyond the standard model
(SM). Theoretically, exclusive semileptonic $B$ decays to $D$ and $D^*$ are well
understood and inclusive semileptonic $B \to X_c\ell\bar\nu$ decay has also been
the focus of extensive research.  Semileptonic $B$ decays to excited charmed
mesons received less attention, but are important for the following reasons.

\begin{enumerate}

\item Recently, \babar, Belle, and LHCb reported discrepancies from the SM
predictions in semi-tauonic decays compared to the $l=e,\, \mu$ light lepton
final states~\cite{Lees:2012xj,Lees:2013uzd, Huschle:2015rga, Aaij:2015yra}. 
Their average shows a disagreement with the SM expectation at the $4 \sigma$
level~\cite{HFAG}.  This tension is intriguing, because it occurs in a
tree-level SM process, and most new physics explanations require new states at
or below 1\,TeV~\cite{Freytsis:2015qca}.

Semileptonic decays into excited charmed mesons with light leptons are an
important background, and their better understanding is needed to improve the
precision of these ratios.

\item Determinations of the CKM matrix element $|V_{cb}|$ from exclusive and
inclusive semileptonic $B$ decays exhibit a nearly $3\sigma$
tension~\cite{HFAG}.  Decays involving heavier charmed mesons are an important
background of untagged exclusive measurements, and are also important in
inclusive $|V_{cb}|$ measurements since efficiency and acceptance effects are
modeled using a mix of exclusive decay modes that includes decays into excited
charmed mesons. 

\item Semi-tauonic decays into excited charmed mesons provide a complementary
probe of the enhancements observed in the semi-tauonic decays to $D$ and $D^*$.
Moreover, the measured semi-tauonic decays to $D$ and $D^*$ appear to saturate
the inclusive $\bar B \to X\, \tau \bar\nu$ rate~\cite{Freytsis:2015qca}. This
motivates measuring this decay, and if the enhancement is verified, new physics
modifying the $D^{(*)}$ rates must also fit the semi-tauonic rates for higher
mass charm states. 

\end{enumerate}

Heavy quark symmetry~\cite{Isgur:1989vq} provides some model independent
predictions for exclusive semileptonic $B$ decays to excited charmed mesons,
even including $\lqcd/m_{c,b}$ corrections~\cite{Leibovich:1997tu}. 
Approximations based on those results constitute the LLSW
model~\cite{Leibovich:1997em}, used in many experimental analyses.  The key
observation was that some of the $\lqcd/m_{c,b}$ corrections to semileptonic
form factors at zero recoil are determined by the masses of orbitally excited
charmed mesons~\cite{Leibovich:1997tu, Leibovich:1997em}.

\begin{table}[b]
\begin{tabular}{ccccc}
\hline\hline
Particle  &    $s_l^{\pi_l}$ &  $J^P$  &  $m$ (MeV)  &  $\Gamma$ (MeV)\\
\hline
$D_0^*$ &  $\frac12^+$  &  $0^+$  &  $2330$  &  270 \\
$D_1^*$ &  $\frac12^+$  &  $1^+$  &  $2427$  &  384 \\[2pt]
\hline
$D_1$ &  $\frac32^+$  &  $1^+$  &  $2421$  &  34 \\
$D_2^*$ &  $\frac32^+$  &  $2^+$  &  $2462$  &  48 \\[2pt]
\hline\hline
$B_1$ &  $\frac32^+$  &  $1^+$  &  $5727$  &  28 \\
$B_2^*$ &  $\frac32^+$  &  $2^+$  &  $5739$  &  23 \\[2pt]
\hline\hline
\end{tabular}
\caption{Isospin averaged masses and widths of some excited $D$ mesons, rounded
to 1\,MeV.  For the $\frac32^+$ states we averaged the PDG with LHCb
measurements~\cite{Aaij:2013sza, Aaij:2015qla} not included in the PDG. The $D_0^*$ mass is
discussed in the text; see Table~\ref{tab:D0data}.  }
\label{tab:charm}
\end{table}

The isospin averaged masses and widths of the four lightest excited $D$ meson
states are shown in Table~\ref{tab:charm}.  In the quark model, they correspond
to combining the heavy quark and light quark spins with $L=1$ orbital angular
momentum.  In the heavy quark limit, the spin-parity of the light degrees of
freedom, $s_l^{\pi_l}$, is a conserved quantum number~\cite{Isgur:1991wq}. 
This spectroscopy is important, because in addition to the impact on the
kinematics, they give important information on heavy quark effective theory (HQET) matrix elements and the QCD
dynamics. The level of agreement between the measurements of the masses and
widths of the excited $D$ states in the top 4 rows of Table~\ref{tab:charm} is
not ideal.  In particular, the mass of the $D_0^*(2400)$ varies in published
papers by 100\,MeV, as shown in Table~\ref{tab:D0data}.  The confidence level of
our mass average in the last row is 5\%.

The masses of a heavy quark spin symmetry doublet of hadrons, $H_\pm$, with
total spin $J_\pm = s_l \pm \frac12$ can be expressed in HQET as
\begin{equation}\label{mass}
m_{H_\pm} = m_Q + \bar\Lambda^H - {\lambda_1^H \over 2 m_Q} 
  \pm {n_\mp\, \lambda_2^H \over 2m_Q} + \ldots \,,
\end{equation}
where $n_\pm = 2J_\pm+1$ is the number of spin states of each hadron,
and the ellipsis denote terms suppressed by more powers of  $\Lambda_{\rm
QCD}/m_Q$.  The parameter $\bar\Lambda^H$ is the energy of the light degrees
of freedom in the $m_Q\to\infty$ limit, and plays an important role, as it is
related to the semileptonic form factors~\cite{Leibovich:1997tu,
Leibovich:1997em}.  We use the notation $\bar\Lambda$, $\bar\Lambda'$, and 
$\bar\Lambda^*$ for the $\frac12^-$, $\frac32^+$, and $\frac12^+$ doublets,
respectively.  The $\lambda_1^H$ and $\lambda_2^H$ parameters are related to the
heavy quark kinetic energy and chromomagnetic energy in hadron $H$.

\begin{table}[t]
\begin{tabular}{ccl}
\hline\hline
m (MeV)  &    $\Gamma$ (MeV)  &  reference  \\
\hline
$2405\pm36$  &  $274\pm45$  &  FOCUS~\cite{Link:2003bd} \\
$2308\pm36$  &  $276\pm66$  &  Belle~\cite{Abe:2003zm} \\
$2297\pm22$  &  $273\pm49$  &  \babar~\cite{Aubert:2009wg} \\
$2360\pm34$  &  $255\pm57$  &  LHCb~\cite{Aaij:2015kqa} \\
\hline
$2330\pm15$  &  $270\pm26$  &  our average \\
\hline\hline
\end{tabular}
\caption{Isospin averaged $D_0^*(2400)$ masses and widths.
The LHCb measurement~\cite{Aaij:2015kqa} is missing from the PDG.}
\label{tab:D0data}
\end{table}

The current data suggest that the $m_{D_1^*} - m_{D_0^*}$ mass splitting is
substantially larger than the $m_{D_2^*} - m_{D_1}$ splitting.  This possibility
was not considered in Refs.~\cite{Leibovich:1997tu, Leibovich:1997em}, since at
that time both of these mass splittings were about 40\,MeV.  The smallness of
$m_{D_2^*} - m_{D_1}$ and $m_{D_1^*} - m_{D_0^*}$ compared to $m_{D^*} - m_D
\simeq 140\,\MeV$ was taken as an indication that the chromomagnetic operator
matrix elements are suppressed for the four $D^{**}$ states, in agreement with
quark model predictions.  We explore the consequences of relaxing this
constraint.

\begin{table}[t]
\begin{tabular}{ccc|cc}
\hline\hline
$s_l^{\pi_l}$ &  Particles	& $\overline{m}$ (MeV) &  Particles & $\overline{m}$ (MeV) \\ \hline 
$\frac12^-$  &  $D$, $D^*$	&  1973  &  $B$, $B^*$  &  5313  \\
$\frac12^+$  &  $D_0^*$, $D_1^*$ &  2403 &  $B_0^*$, $B_1^*$ & ---\\
$\frac32^+$  &  $D_1$, $D_2^*$	&  2445  &  $B_1$, $B_2^*$  & 5734 \\
\hline\hline
\end{tabular}
\caption{Isospin and heavy quark spin symmetry averaged masses of lightest $B$
and $D$ multiplets (with weights $2J+1$).}
\label{tab:averagemass}
\end{table}

The isospin and heavy quark spin symmetry averaged masses in
Table~\ref{tab:averagemass} and Eq.~(1.10) in Ref.~\cite{Leibovich:1997em},
which is valid to ${\cal O} (\lqcd^3/m_{c,b}^2)$, yield $\bar\Lambda' -
\bar\Lambda = 0.40\,\GeV$ (using $m_b=4.8\,\GeV$ and $m_c=1.4\,\GeV$, but the
sensitivity to this is small).  While the masses of the broad $D_0^*$ and
$D_1^*$ states changed substantially since the 1990s, their $2J+1$ weighted
average mass is essentially unchanged compared to Ref.~\cite{Leibovich:1997em}. 
We estimate $\bar\Lambda' - \bar\Lambda^* \simeq 0.04\,\GeV$ from
Table~\ref{tab:averagemass}.  We summarize the parameters used in
Table~\ref{tab:input_summary}.  The uncertainty of $\bar\Lambda$ is
substantially greater than that of $\bar\Lambda' - \bar\Lambda$ and
$\bar\Lambda' - \bar\Lambda^*$, but as we see below, our results are less
sensitive to $\bar\Lambda$ than to these differences.  The parameters with $s$
subscripts in Table~\ref{tab:input_summary} are relevant for $B_s\to
D_s^{**}\ell\bar\nu$ discussed in Sec.~\ref{sec:SMBs_rate}.

\begin{table}[t]
\begin{tabular}{c|ccc|ccc}
\hline\hline
Parameter  &  $\bar\Lambda$  &  $\bar\Lambda'$  &  $\bar \Lambda^*$ 
  &  $\bar\Lambda_s$  &  $\bar\Lambda'_{s}$  &  $\bar \Lambda^*_{s}$ \\
Value [GeV]  &  0.40 & 0.80 & 0.76  &  0.49 & 0.90 & 0.77 \\
\hline\hline
\end{tabular}
\caption{The HQET parameter estimates used.}
\label{tab:input_summary}
\end{table}

Another effect suppressed in the heavy quark limit and neglected in
Refs.~\cite{Leibovich:1997tu, Leibovich:1997em}, is the mixing between $D_1$ and
$D_1^*$.  It was recently argued that this could be
substantial~\cite{Klein:2015doa}.  However, even a small mixing of the $D_1$
with the much broader $D_1^*$ would yield $\Gamma_{D_1} > \Gamma_{D_2^*}$, in
contradiction with the data, which suggests that this $\lqcd/m_c$ effect may
be small~\cite{Lu:1991px, Kilian:1992hq, Falk:1995th}.  Until the masses are
unambiguously measured, we neglect the effects of this mixing, which we expect
to be modest, and leave it for another study, should future data suggest that it
is important.

The rest of this paper is organized as follows.  Section~\ref{sec:SMrate}
reviews the $B \to D^{**}\, \ell\, \bar\nu$ decays to the four states
collectively denoted
\beq
D^{**} = \{D_0^*,\, D_1^*,\, D_1,\, D_2^*\}\,,
\eeq
and provides expressions for these decay rates with the full lepton mass
dependence.  In Sec.~\ref{sec:FF} the expansion of the form factors based on
heavy quark symmetry~\cite{Leibovich:1997tu, Leibovich:1997em} is briefly
reviewed. Section~\ref{sec:FF_Fit} summarizes the experimental analysis to
determine the leading Isgur-Wise function normalization and slope, and we obtain
predictions for the ratios of semileptonic rates for $\tau$ and light leptons,
\beq\label{Rdef}
R({D^{**}}) = \frac{ \mathcal{B}( B \to D^{**} \tau\, \bar\nu)}
   {\mathcal{B}( B \to D^{**} l\, \bar\nu)}\,, \qquad l=e,\mu\,.
\eeq
Section~\ref{sec:SMBs_rate} studies predictions for $B_s\to
D_s^{**}\ell\bar\nu$.  Section~\ref{sec:2HDM} explores extensions of the SM with
scalar currents.  Predictions for the rates and $R({D^{**}})$ are derived to
illustrate the complementary sensitivity of each mode. Section~\ref{sec:summary}
summarizes our main findings.

\section{The $B \to D^{**} \ell\, \bar\nu$ decays in the SM}
\label{sec:SMrate}

The effective SM Lagrangian describing $b \to c\, \ell\, \bar\nu$ is
\beq
\mathcal{L}_{\rm eff} = - \frac{4 G_F} { \sqrt{2}}\, V_{cb}\, 
  \big( \bar c\, \gamma_\mu P_L b \big) \big( \bar\nu\, \gamma^\mu P_L \ell 
  \big) + \text{h.c.}\,,
\eeq
with the projection operator $P_L = (1 - \gamma_5) / 2$, $G_F$ is the Fermi
constant, and $\ell$ denotes any one of $e,\mu,\tau$. The matrix elements of the
$B\to D^{**}$ vector and axial-vector currents ($V^\mu=\bar c\,\gamma^\mu\,b$
and $A^\mu=\bar c\,\gamma^\mu\gamma_5\,b$) can be parameterized for the
$\frac32^+$ states as
\begin{eqnarray}\label{formf32}
{\langle D_1(v',\epsilon)| V^\mu |B(v)\rangle \over \sqrt{m_{D_1}\,m_B}}
  &=& f_{V_1} \epsilon^{* \mu} 
  + (f_{V_2} v^\mu + f_{V_3} v'^\mu) (\epsilon^*\!\cdot v) \,, \nn\\
{\langle D_1(v',\epsilon)| A^\mu |B(v)\rangle \over \sqrt{m_{D_1}\,m_B}}
  &=& i\, f_A\, \varepsilon^{\mu\alpha\beta\gamma} 
  \epsilon^*_\alpha v_\beta v'_\gamma \,, \nn\\
{\langle D^*_2(v',\epsilon)| A^\mu |B(v)\rangle \over\sqrt{m_{D_2^*}\,m_B}}
  &=& k_{A_1}\, \epsilon^{*\mu\alpha} v_\alpha \nn\\
  &&{} + (k_{A_2} v^\mu + k_{A_3} v'^\mu)\,
  \epsilon^*_{\alpha\beta}\, v^\alpha v^\beta \,, \nn\\
{\langle D^*_2(v',\epsilon)| V^\mu |B(v)\rangle \over\sqrt{m_{D_2^*}\,m_B}}
  &=& i\,k_V\, \varepsilon^{\mu\alpha\beta\gamma} 
  \epsilon^*_{\alpha\sigma} v^\sigma v_\beta v'_\gamma \,,
\end{eqnarray}
while for the $\frac12^+$ states
\begin{eqnarray}\label{formf12}
\langle D_0^*(v')| V^\mu |B(v)\rangle &=& 0, \nn\\
{\langle D_0^*(v')| A^\mu |B(v)\rangle \over\sqrt{m_{D_0^*}\,m_B}}
  &=& g_+\, (v^\mu+v'^\mu) + g_-\, (v^\mu-v'^\mu) \,, \nn\\
{\langle D_1^*(v',\epsilon)| V^\mu |B(v)\rangle \over\sqrt{m_{D_1^*}\,m_B}}
  &=& g_{V_1} \epsilon^{* \mu} 
  + (g_{V_2} v^\mu + g_{V_3} v'^\mu)\, (\epsilon^*\!\cdot v) \,, \nn\\
{\langle D_1^*(v',\epsilon)| A^\mu |B(v)\rangle \over\sqrt{m_{D_1^*}\,m_B}}
  &=& i\, g_A\, \varepsilon^{\mu\alpha\beta\gamma}\, 
  \epsilon^*_\alpha v_\beta\, v'_\gamma \,.
\end{eqnarray}
Here the form factors $g_i$, $f_i$ and $k_i$ are dimensionless functions of
$w = v \cdot v'$.  At zero recoil ($w=1$ and $v=v'$) only the $g_+$, $g_{V_1}$,
and $f_{V_1}$ form factors can contribute, since $v'$ dotted into the
polarization ($\epsilon^{*\mu}$ or $\epsilon^{*\mu\alpha}$) vanishes.  The
variable $w$ is related to the four-momentum transfer squared, $q^2 = (p_B -
p_{D^{**}})^2$, as
\beq
w = v \cdot v' = \frac{m_B^2 + m_{D^{**}}^2 - q^2 }{2\, m_B\, m_{D^{**}}} \,.
\eeq

\subsection{Differential decay rates}

We define $\theta$ as the angle between the charged lepton and the charmed meson
in the rest frame of the virtual $W$ boson, i.e., in the center of momentum
frame of the lepton pair.  It is related to the charged lepton energy via
\begin{align}\label{yct}
y &= 1 - rw - r\sqrt{w^2-1}\, \cos\theta \nn\\
  & + \rl\, \frac{1 - rw+ r\sqrt{w^2-1}\, \cos\theta}{1-2 r w+r^2}\,,
\end{align}
where $y=2E_\ell/m_B$ is the rescaled lepton energy and $\rl = m_\ell^2/m_B^2$.
For the double differential rates in the SM for the $s_l^\pi = \frac32^+$
states we obtain
\begin{widetext}
\begin{align}\label{D1rate}
& \frac{\d\Gamma_{D_1}}{\d w\, \d \cos\theta} =
  3 \Gamma_0\, r^3 \sqrt{w^2-1}\, \big(1+r^2-\rl-2 r w\big)^2 \\*
&\times \Bigg\{\! \sin^2\theta\, \bigg[
  \frac{\big[f_{V_1} (w-r) + (f_{V_3} + r f_{V_2})(w^2-1)\big]^2}{(1+r^2-2rw)^2} 
  + \rl \frac{f_{V_1}^2 + \big(2f_A^2+f_{V_2}^2+f_{V_3}^2 + 2f_{V_1}f_{V_2}+
   2wf_{V_2}f_{V_3}\big) (w^2-1)}{2 (1+r^2-2rw)^2}\bigg] \nn\\*
&\qquad + (1+\cos^2\theta)\, \bigg[ \frac{f_{V_1}^2 +f_A^2(w^2-1)}{1+r^2-2r w}
  + \rl \frac{ [f_{V_1}^2 + (w^2-1)f_{V_3}^2](2w^2-1+r^2-2rw)}
  {2 (1+ r^2-2 rw)^3}  \nn\\*
&\qquad\quad + \rl (w^2-1) \frac{2f_{V_1} f_{V_2}(1-r^2) + 4 f_{V_1}f_{V_3}(w-r)   
  + f_{V_2}^2(1-2rw-r^2+2r^2w^2) 
  + 2f_{V_2}f_{V_3} (w-2r+r^2w)}{2 (1+ r^2-2 rw)^3} \bigg] \nn\\*
&\qquad - 2 \cos\theta\,\sqrt{w^2-1}\, \bigg[
  \frac{2 f_A f_{V_1}}{1+r^2-2 r w}
  - \rl \frac{\big[f_{V_1} (w-r) + (f_{V_3} + r f_{V_2}) (w^2-1)\big] 
  [f_{V_1} +f_{V_2}(1-rw) +f_{V_3}(w-r)]}{(1+r^2-2r w)^3}\bigg]\! \Bigg\}\,, \nn
\end{align}
where $r=m_{D^{**}}/m_B$ for each $D^{**}$ state, as
appropriate, $\Gamma_0 = {G_F^2\,|V_{cb}|^2\,m_B^5 /(192\pi^3)}$.
For $B\to D_2^*\ell\bar\nu$ we find
\begin{align}\label{D2rate}
& \frac{\d\Gamma_{D_2^*}}{\d w\, \d \cos\theta} =
  \Gamma_0\, r^3 (w^2-1)^{3/2}\, \big(1+r^2-\rl-2 r w\big)^2 \\*
&\times\! \Bigg\{\! \sin^2\theta\, \bigg[
  \frac{2\big[k_{A_1}(w-r) + (k_{A_3} + r k_{A_2})(w^2-1)\big]^2}{(1+r^2-2rw)^2} 
  + \rl \frac{3k_{A_1}^2 +\big(3 k_V^2 +2k_{A_2}^2 +2k_{A_3}^2 + 4k_{A_1}k_{A_2}
  + 4w k_{A_2}k_{A_3}\big) (w^2-1)}{2 (1+r^2-2 r w)^2} \bigg] \nn\\*
&\qquad + (1+\cos^2\theta)\, \bigg[ 
  \frac32\, \frac{k_{A_1}^2 + k_V^2 (w^2-1)}{1+r^2-2 r w} 
  + \rl \frac{ [k_{A_1}^2 + (w^2-1)k_{A_3}^2](2w^2-1+r^2-2rw)}{(1+r^2-2 rw)^3} 
  \nn\\*
&\qquad\quad + \rl (w^2-1) \frac{2k_{A_1} k_{A_2}(1-r^2) + 4 k_{A_1}k_{A_3}(w-r)   
   + k_{A_2}^2(1-2rw-r^2+2r^2w^2) 
   + 2k_{A_2}k_{A_3} (w-2r+r^2w)}{(1+ r^2-2 rw)^3} \bigg] \nn\\*
&\qquad - 2\cos\theta\,\sqrt{w^2-1}\,
  \bigg[ \frac{3 k_V k_{A_1}}{1+r^2-2 r w}
  - 2 \rl \frac{\big[k_{A_1} (w-r)+(k_{A_3} + r k_{A_2}) (w^2-1)\big] 
  [k_{A_1} + k_{A_2}(1-rw)+ k_{A_3}(w-r)]}{(1+r^2-2r w)^3} \bigg]\!\Bigg\}. \nn
\end{align}

For the $\frac12^+$ $D^{**}$ mesons, the rate for $\d\Gamma_{D_1^*}/\d w\, \d
\cos\theta$ is obtained from the $D_1$ rate above via the replacements $\{f_A\to
g_A,\, f_{V_1}\to g_{V_1},\, f_{V_2}\to g_{V_2},\, f_{V_3}\to g_{V_3}\}$, and
for $B\to D_0^*\ell\bar\nu$ we find
\begin{align}\label{D0rate}
\frac{\d\Gamma_{D_0^*}}{\d w\, \d \cos\theta} & = 
  3 \Gamma_0\, r^3 \sqrt{w^2-1} \big(1-2 r w+r^2-\rl\big)^2 \bigg\{\!
  \sin^2\theta\, \frac{ [g_+(1+r) - g_-(1-r)]^2\, (w^2-1) 
  + \rl [ g_+^2(w+1) + g_-^2 (w-1)]}{(1+r^2-2 rw)^2} \nn\\*
&\quad + (1+\cos^2\theta)\, \rl\, \frac{
  \big[g_+^2 (w+1)+g_-^2 (w-1)\big] \big(w-2r+r^2 w\big)
  - 2 g_- g_+ (1-r^2) (w^2-1)}{(1+r^2-2 r w)^3} \nn\\*
&\quad - 2 \cos\theta\, \rl\, \sqrt{w^2-1}\, \frac{[g_+ (1+r) - g_- (1-r)]\,
  [g_- (1+r) (w-1) - g_+ (1-r) (w+1)]}{(1+r^2-2 r w)^3} \bigg\} \,.
\end{align}
\end{widetext}
The $\sin^2\theta$ terms are the helicity zero rates, while the $1+\cos^2\theta$
and $\cos\theta$ terms determine the helicity $\lambda=\pm1$ rates.  The decay
rates for $|\lambda|=1$ vanish for massless leptons at maximal recoil, $w_{\rm
max} = (1+r^2-\rt)/(2r)$, as implied by the $(1-2rw+r^2-\rt)$ factors.

At zero recoil, the leading contributions to the matrix elements of the weak
currents are determined by $f_{V_1}(1)$, $g_{V_1}(1)$, and $g_+(1)$, which are
of order $\Lambda_{\rm QCD}/m_{c,b}$.  The contributions of other form factors
are suppressed by powers of $w-1$.  The model independent result is that these
numerically significant ${\cal O}(\Lambda_{\rm QCD}/m_{c,b})$ effects at $w=1$
are determined in terms of hadron masses and the leading Isgur-Wise function,
without dependence on any subleading ${\cal O}(\lqcd/m_{c,b})$ Isgur-Wise
functions~\cite{Leibovich:1997tu}. The results in
Eqs.~(\ref{D1rate})--(\ref{D0rate}) show that this holds even for $\rl \neq 0$,
and treating $\rl = {\cal O}(1)$, since
\begin{eqnarray}
\sqrt6\, f_{V_1}(w) &=& - \big[w^2-1
  + 8\, \varepsilon_c (\bar\Lambda'-\bar\Lambda)\big] \tau(w)
  + \ldots \,, \nn\\*
g_+(w) &=& -\frac32\, (\varepsilon_c+\varepsilon_b)\, 
  (\bar\Lambda^*-\bar\Lambda)\, \zeta(w) + \ldots \,, \\*
g_{V_1}(w) &=& \big[w-1
  + (\varepsilon_c-3\,\varepsilon_b)\, 
  (\bar\Lambda^*-\bar\Lambda) \big] \zeta(w) + \ldots \,,\nn
\end{eqnarray}
where $\varepsilon_{c,b} = 1/2m_{c,b}$ and the ellipses denote ${\cal
O}[\varepsilon_{c,b} (w-1)]$, ${\cal O}[(w-1)\, \alpha_s]$, and higher order
terms.  In contrast, Eqs.~(\ref{FFD0}) -- (\ref{FFD2}) in
Appendix~\ref{App:formfactors} show that the other form factors depend on
subleading Isgur-Wise functions, even at $w=1$.  The $B\to D^{**}\tau\bar\nu$
rate and $R(D^{**})$ were previously studied using QCD sum rule calculation of
the leading Isgur-Wise function~\cite{Biancofiore:2013ki}.

\subsection{Form factors and approximations}
\label{sec:FF}

Heavy quark symmetry~\cite{Isgur:1989vq} implies that in the $m_{c,b} \gg
\Lambda_{\rm QCD}$ limit the form factors defined in Eqs.~(\ref{formf32}) and
(\ref{formf12}) are determined by a single universal Isgur-Wise function, which
we denote by $\tau(w)$ and $\zeta(w)$, respectively, for the $\frac32^+$ and
$\frac12^+$ states.\footnote{Another often used notation in the literature is
$\tau(w) = \sqrt3\, \tau_{3/2}(w)$ and $\zeta(w) = 2\, \tau_{1/2}(w)$.}
In the $m_{c,b} \gg \lqcd$ limit, the contributions of $\tau$ and $\zeta$ vanish
at $w=1$, thus the rates near zero recoil entirely come from $\lqcd / m_{c,b}$
corrections.  Some of the $\lqcd / m_{c,b}$ corrections can be expressed in
terms of the leading Isgur-Wise function and meson mass
splittings~\cite{Leibovich:1997tu, Leibovich:1997em}.  The full expressions are
reproduced for completeness in Appendix~\ref{App:formfactors}.  The leading
order Isgur-Wise function for the $\frac32^+$ states can be parametrized as
\beq \label{eq:leading_IW}
 \tau(w) = \tau(1) \big[ 1 + (w - 1)\, \tau'(1) +\ldots \big] \,,
\eeq
and $\tau(1)$ can be constrained from the measured $\bar B \to D_1\, \ell\,
\bar\nu$ branching fraction.  In Ref.~\cite{Leibovich:1997em} the dependence of
the predictions was studied as a function of $\tau'$, taken to be near $-1.5$,
based on model predictions~\cite{ISGW, Colangelo:1992kc, Veseli:1996kn,
Morenas:1997nk}; with more data a fit to all information is preferred.

In any nonrelativistic constituent quark model with spin-orbit independent
potential~\cite{Isgur:1990jf, Veseli:1996kn} the Isgur-Wise functions for the
$s_l^\pi = \frac32^+$ and  $s_l^\pi = \frac12^+$ states are related,
\beq\label{eq:quark_model_IW}
 \zeta(w) = \frac{w+1}{\sqrt{3}} \, \tau(w) \,.
\eeq
This relation determines the form factor for the broad states from the narrow
states' form factor slope and normalization.  (See Refs.~\cite{Blossier:2009vy,
Atoui:2013ksa} for exploratory calculations of these Isgur-Wise functions using
lattice QCD.)

The form factors at order $\lqcd/m_{c,b}$ depend on several additional
functions. The $\tau_i$ and $\zeta_i$ parameterize corrections to the $b\to c$
current, while $\eta_i$ and $\chi_i$ parameterize matrix elements involving time
ordered products of subleading terms in the HQET Lagrangian.  Since the range in
$w$ is small, for simplicity these functions may be taken to be proportional to
the leading Isgur-Wise function.  
Since the kinetic energy operator does not violate heavy quark spin symmetry,
its effects can be absorbed into the leading Isgur-Wise functions by the
replacements $\tau \to \tau+\varepsilon_c\, \eta_{\rm ke}^{(c)}+\varepsilon_b\,
\eta_{\rm ke}^{(b)}$ and $\zeta \to \zeta + \varepsilon_c\, \chi_{\rm ke}^{(c)}
+ \varepsilon_b\, \chi_{\rm ke}^{(b)}$.

In what Ref.~\cite{Leibovich:1997em} called \ApproxA, the kinematic range, $0
\leq w-1 \lesssim 1.3$, is treated as a quantity of order $\lqcd/m_{c,b}$, and
the rates are expanded to order $\varepsilon^2$ beyond the $\sqrt{w^2-1}$ phase
space factors, where $\varepsilon = {\cal O}(w-1) = {\cal O}(\lqcd/m_{c,b})$. 
Its generalization for $\rho_\ell \neq 0$ is given in
Appendix~\ref{sec:ApproxA}.  An advantage is that this approach unambiguously
truncates the number of fit parameters to a small number; only 5 parameters
occur for each of the the $\frac32^+$ and $\frac12^+$ states, $\{\tau,\,
\hat\tau',\, \hat\eta_1,\, \hat\eta_3,\, \hat\eta_b\}$ and $\{\zeta,\,
\hat\zeta',\, \hat\chi_1,\, \hat\chi_2,\, \hat\chi_b\}$, respectively.  Among
these, the first two are the zero-recoil values and slopes of the Isgur-Wise
functions, and the latter three are matrix elements of time ordered products
involving the chromomagnetic operator.  These $\eta$'s and $\chi$'s were
neglected in Ref.~\cite{Leibovich:1997em}.

To study lepton universality, another reason to consider \ApproxA is
because it would be advantageous both theoretically~\cite{Freytsis:2015qca} and
experimentally~\cite{Bernlochner:2015mya} to consider instead of $R(X)$ in
Eq.~(\ref{Rdef}), ratios in which the range of $q^2$ integration is the same in
the numerator and the denominator,
\beq
\widetilde R(X) = \frac{\ds \int_{m_\tau^2}^{(m_B-m_X)^2} 
  \frac{\d\Gamma(B \to X \tau\bar\nu)}{\d q^2}\, \d q^2 }
  {\ds \int_{m_\tau^2}^{(m_B-m_X)^2} 
  \frac{\d\Gamma(B \to X l\bar\nu)}{\d q^2}\, \d q^2} \,.
\eeq
Including the $0 < q^2 < m_\tau^2$ region in the denominator in Eq.~(\ref{Rdef})
dilutes the sensitivity to new physics, and the uncertainties of the form
factors increase at larger $w$ (smaller $q^2$).  Taking the average $D^{**}$
mass as near 2.4\,GeV, the kinematic range in $B\to D^{**}\tau\bar\nu$ is only
about $1 \leq w \lesssim 1.2$.  \ApproxA should work better for this reduced
kinematic range, $0 \leq w-1 \lesssim 0.2$, than for the total $D^{**}$ rates.

In \ApproxB and C the full $w$ dependence known at order $\lqcd/m_{c,b}$ is included. 
To reduce the number of free parameters, Ref.~\cite{Leibovich:1997em} assumed a
linear shape for the leading Isgur-Wise functions, and that the subleading ones
have the same shapes.  Motivated by the form of the constraints imposed by the
equations of motions on the $\lqcd/m_{c,b}$ corrections, two variants were
explored,
\begin{align}
\mbox{Approx.~B}_1: \ &  
  \begin{cases}
  \frac32^+ \mbox{ states: } \tau_1 = \tau_2 = 0\,, \\*
  \frac12^+ \mbox{ states: } \zeta_1 = 0\,,
  \end{cases}\\*
\mbox{Approx.~B}_2: \ &  
  \begin{cases}
  \frac32^+ \mbox{ states: } 
    \tau_1 = \bar\Lambda\tau,\ \tau_2 = -\bar\Lambda'\tau\,,\hspace*{-1mm} \\*
  \frac12^+ \mbox{ states: } \zeta_1 = \bar\Lambda\zeta\,.
  \end{cases}
\end{align}
In this paper we also study a generalization,
\begin{align}
\mbox{Approx.~C}: \ &  
  \begin{cases}
  \frac32^+ \mbox{ states: } 
    \tau_1 =  \hat \tau_1 \tau,\ \tau_2 = \hat\tau_2 \tau\,,\hspace*{-1mm} \\*
  \frac12^+ \mbox{ states: } \zeta_1 = \hat \zeta_1 \zeta\,,
  \end{cases}
\end{align}
where the normalization of the subleading Isgur-Wise functions is determined
from experimental constraints. We also study in \ApproxC the impact of not
neglecting the chromomagnetic matrix elements.  As explained above, this is
motivated by the sizable mass splitting, $m_{D_1^*} - m_{D_0^*}$.  Note also the
large coefficients of $\eta_1$ (10 and 12) in the $f_{V_2}$ and $f_{V_3}$ form
factors in Eq.~(\ref{FFD1}).

\section{Form Factor Fit}
\label{sec:FF_Fit}

The parameters that occur in the expansions of the form factors can be
constrained by the measured semileptonic rates.  Belle and \babar measured the
total branching fraction of the four $D^{**}$ states and Belle in addition the
$q^2$ distribution of $B\to D_2^*l\bar\nu$ and $B \to D_0
l\bar\nu$~\cite{Liventsev:2007rb, Aubert:2008ea}. The measurements were carried
out in the $D^{**} \to D^{(*)+}\, \pi^-$ channels, and to confront the measured
branching fractions with decay rate predictions, one needs to account for
missing isospin conjugate decay modes and other missing contributions. The
missing isospin modes can be accounted for with the factor
\begin{equation}
f_\pi = \frac{\mathcal{B}(D^{**} \to D^{(*)\,0} \, \pi^- )}{ \mathcal{B}(D^{**} \to D^{(*)} \pi)}  = \frac{2}{3} \, .  \
\end{equation}
The measurements of the $B^- \to D_2^*{}^0\, l\, \bar\nu$ branching fraction
that enter the world average are converted to only account for the $D_2^*{}^0
\to D^{*\, +}\, \pi^-$ decay. To account for the missing $D_2^*{}^0 \to D^{+} \,
\pi^-$ decay a correction factor
\begin{equation}
f_{D_2^*} = \frac{\mathcal{B}(D_2^*{}^0 \to D^{*\,+}\, \pi^-)}{\mathcal{B}(D_2^*{}^0 \to D^+ \, \pi^-) } = 0.65 \pm 0.06 \, ,
\end{equation}
from Ref.~\cite{PDG} is applied.

The measurements of the $B^- \to D_1^0\, l\,\bar\nu$ branching fraction do not
include contributions of the observed three-body decay of the $D_1$. This
is corrected with a factor
\begin{equation}
f_{D_1} = \frac{\mathcal{B}(D_1^0 \to D^{*\,+}\, \pi^-)}
  {\mathcal{B}(D_1^0 \to D^0 \, \pi^+ \, \pi^-) } = 2.32 \pm 0.54 \,,
\end{equation}
as calculated from the ratio of nonleptonic $B^+ \to \bar D_1^0\, \pi^+$ decays
of Ref.~\cite{Aaij:2011rj}. Assuming no intermediate resonances are present in
the three-body decay of a $D^{**}$ meson, one obtains an isospin correction
factor of
\begin{equation}
f_{\pi\pi} = \frac{\mathcal{B}(D^{**} \to D^{(*)-} \, \pi^+ \, \pi^- )}{ \mathcal{B}(D^{**} \to D^{(*)} \pi \pi)}  = \frac{9}{16} \, .  \
\end{equation}
If the three-body final state of a $D^{**}$ meson is reached through resonances,
i.e., via $f_0(500) \to \pi \pi$ or $\rho \to \pi \pi$ decays, this factor is
either $2/3$ or $1/3$, respectively. In what follows we adapt the
prescription proposed in Ref.~\cite{Lees:2015eya} and apply an average
correction factor
\begin{equation}
 f_{\pi\pi} = \frac{1}{2} \pm \frac{1}{6} \, ,
\end{equation}
with an uncertainty spanning all three scenarios. After these
corrections we make the explicit assumption that
\begin{align}
\mathcal{B}(\bar D_2^* \to \bar  D \, \pi) + \mathcal{B}(\bar  D_2^* \to \bar  D^* \, \pi) \,& = 1\,, \nn \\
\mathcal{B}(\bar D_1 \to \bar D^* \, \pi) + \mathcal{B}(\bar D_1 \to \bar D \, \pi \, \pi) \,& = 1\,, \nn \\
\mathcal{B}(\bar D_1^* \to \bar D^* \, \pi) \, &= 1\,, \nn \\
 \mathcal{B}(\bar D_0^* \to\bar  D \, \pi) \, & = 1\,,  \label{eq:bf_assumptions}
\end{align}
and then all semileptonic rates and differential rates can be related.
Table~\ref{tab:sl_input} summarizes the corrected total branching fractions. The
summed $B \to D^{(*)} \pi \pi\, l \, \bar \nu_\ell$ contributions can be
compared with the measurement of Ref.~\cite{Lees:2015eya}. The reported
semi-inclusive $B^+ \to D \, \pi \, \pi\, l \, \bar \nu_\ell$ rates can be
nearly accommodated by the expected $D_1 \to D \, \pi \, \pi$ contribution
\begin{align}
 \mathcal{B}(B^+ \to \bar D^0 \, \pi \, \pi \, l\, \bar\nu ) - \mathcal{B}(B^+ \to \bar D_1^0 ( \to \bar D^0 \, \pi \, \pi) \, l\, \bar\nu )
 \nn \\*
 = \left( 0.06 \pm 0.16 \right) \times 10^{-2} \, .
\end{align}
Decays of the type $\bar D^{**} \to \bar D^* \pi \pi$ have been searched
for~\cite{Abe:2004sm}, but no sizable contribution that could explain the large
reported $\mathcal{B}(B^+ \to \bar D^{* \, 0} \, \pi \, \pi \, l\, \bar\nu )$
branching fraction~\cite{Lees:2015eya} have been observed. It seems
likely that such contributions originate either from higher excitations or
nonresonant semileptonic decays, which would not affect the predictions
discussed in this paper. Table~\ref{tab:sl_input_diff} summarizes the
measured normalized differential decay rates of $B^+ \to \bar D_2^*{}^0 \, l\,
\bar\nu$ and $B^+ \to \bar D_0^*{}^0 \, l\, \bar\nu$ as functions of $w$.

\begin{table}[tb]
\begin{tabular}{ll}
\hline\hline
Decay mode & Branching fraction \\
\hline
$B^+ \to \bar D_2^*{}^0 \, l\, \bar\nu$  & $( 0.30 \pm 0.04) \times 10^{-2}$ \\ 
$B^+ \to \bar D_1^0 \, l\, \bar\nu$  & $( 0.67 \pm 0.05) \times 10^{-2}$ \\ 
$B^+ \to \bar D_1^*{}^0 \, l\, \bar\nu$  & $( 0.20 \pm 0.05) \times 10^{-2}$ \\ 
$B^+ \to \bar D_0^*{}^0 \, l\, \bar\nu$  & $( 0.44 \pm 0.08) \times 10^{-2}$ \\ 
\hline\hline
\end{tabular}
\caption{The corrected world averages of the semileptonic decay rates into
excited charmed mesons~\cite{PDG}. The corrections described in the text involve
factors to account for missing isospin conjugate modes and observed three-body
decays.}
\label{tab:sl_input}
\end{table}

\begin{table}[tb]
\begin{tabular}{ccc}
\hline\hline
 $w$ & $B^+ \to \bar D_2^*{}^0 \, l\, \bar\nu$ & $B^+ \to \bar D_0^*{}^0 \, l\, \bar\nu$  \\ \hline
$1.00-1.08$ & $0.06 \pm 0.02$ & $0.05 \pm 0.02$ \\
$1.08-1.16$ & $0.30 \pm 0.05$ & $0.02 \pm 0.05$ \\
$1.16-1.24$ & $0.38 \pm 0.03$ & $0.30 \pm 0.08$ \\
$1.24-1.32$ & $0.26 \pm 0.06$ & $0.30 \pm 0.09$ \\
$1.32-1.40$ & --- & $0.33 \pm 0.13$ \\
\hline\hline
\\
\end{tabular}
\caption{The normalized differential decay rates for $B^+ \to \bar D_2^*{}^0\,
l\, \bar\nu$ and $B^+ \to \bar D_0^*{}^0\, l\, \bar\nu$ as functions of
$w$~\cite{Liventsev:2007rb}.}
\label{tab:sl_input_diff}
\end{table}

Additional constraints on the form factors at maximal recoil come from
nonleptonic $B^0 \to D^{**\,-} \, \pi^+$ decays. Factorization should be a good
approximation for $B$ decays into charmed mesons and a charged
pion~\cite{Neubert:1997hk, Bauer:2001cu}. Contributions that violate
factorization are suppressed by $\Lambda_{\rm QCD}$ divided by the energy of the
pion in the $B$ restframe or by $\alpha_s(m_Q)$. Neglecting the pion mass, the
two-body decay rate, $\Gamma_\pi$, is related to the differential decay rate
$\text{d}\Gamma_{\rm sl} / \text{d} w$ at maximal recoil for the analogous
semileptonic decay (with the $\pi$ replaced by the $l \bar \nu$ pair)
\begin{align}
 \Gamma_\pi = \frac{3 \pi^2 \left| V_{ud} \right|^2 C^2 f_\pi^2}{m_B^2\, r}
 \left( \frac{\d \Gamma_{\rm sl} }{ \d w} \right)_{w_{\rm max}} \,.
\end{align}
Here $C$ is a combination of Wilson coefficients of four-quark operators and
numerically $\left| V_{ud} \right| C$ is very close to unity.
Table~\ref{tab:nl_input} summarizes the measured nonleptonic rates, after all
correction factors for missing isospin and three-body decays are applied. The
smallness of ${\cal B}(B^0 \to D_0^*{}^- \pi^+)$ is
puzzling~\cite{Jugeau:2005yr, Bigi:2007qp}, and measurements using the full
\babar and Belle data sets would be worthwhile. It would also be interesting to
measure in Belle~II the color suppressed $B^0 \to D^{**\,0}\pi^0$ rates, for
which soft collinear effective theory (SCET) predicts ${\cal B}(B^0 \to D_2^{*\,0}\pi^0) / {\cal B}(B^0 \to
D_1^0\pi^0) =1$~\cite{Mantry:2004pg}.

The narrow and broad states semileptonic and narrow states nonleptonic inputs are analyzed to construct a likelihood
to determine the form factor parameters of \ApproxA, B and C. This is
done separately for the narrow $\frac32^+$ and broad $\frac12^+$ states.

\begin{table}[tb]
\begin{tabular}{ll}
\hline\hline
Decay mode  &  Branching fraction \\
\hline
$B^0 \to D_2^*{}^- \pi^+$		& $\left( 0.59 \pm  0.13 \right) \times 10^{-3}$ \\
$B^0 \to D_1^-  \pi^+$  & $\left( 0.75 \pm 0.16 \right) \times 10^{-3}$ \\
$B^0 \to D_0^*{}^- \pi^+$  & $\left( 0.09 \pm 0.05 \right) \times 10^{-3}$ \\
\hline\hline
\end{tabular}
\caption{World averages of nonleptonic $B^0\to D^{**\,-}\pi^+$ branching
ratios~\cite{PDG}, after the corrections described in the text are applied.}
\label{tab:nl_input}
\end{table}

\subsection{\ApproxA}

The main parameters that determine \ApproxA are the normalization and slope of
the leading Isgur-Wise function for the narrow and broad states, $\{\tau(1)$,
$\tau'\}$ and $\{\zeta(1)$, $\zeta'\}$. In addition, the inclusion of one or two
subleading Isgur-Wise functions parameterizing chromomagnetic contributions is
explored. These are extracted by building a likelihood using experimental
quantities, which are less sensitive to the absence of subleading Isgur-Wise
functions from matrix elements of subleading currents in \ApproxA (see,
Appendix~\ref{sec:ApproxA}).  These are the semileptonic branching fractions to
the narrow $\frac32^+$ states and the nonleptonic $B^0 \to D_2^*{}^- \pi^+$ 
branching fraction. The constraint from the nonleptonic $B^0 \to D_1^- \pi^+$
branching fraction is not included in the fit, as the semileptonic rate to
$D_1$ near $q^2 = m_\pi^2$ receives large corrections from subleading Isgur-Wise
functions that do not enter \ApproxA. Such contributions only mildly affect the
total branching fraction. The analysis of the broad $\frac12^+$ states uses the
measured semileptonic branching fractions only. 

\begin{figure*}[th]
\centerline{
\includegraphics[width=0.5\textwidth]{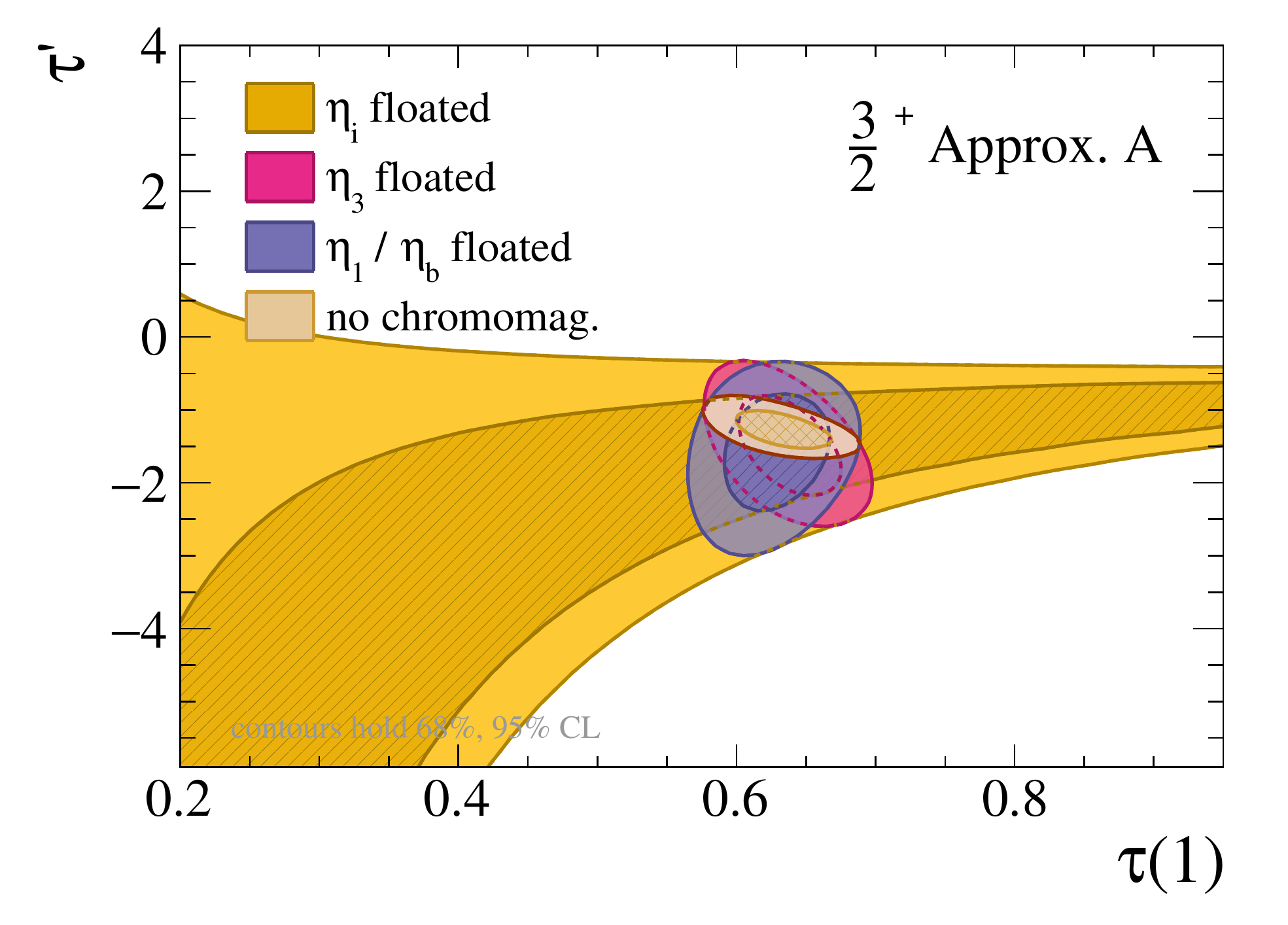}
\includegraphics[width=0.5\textwidth]{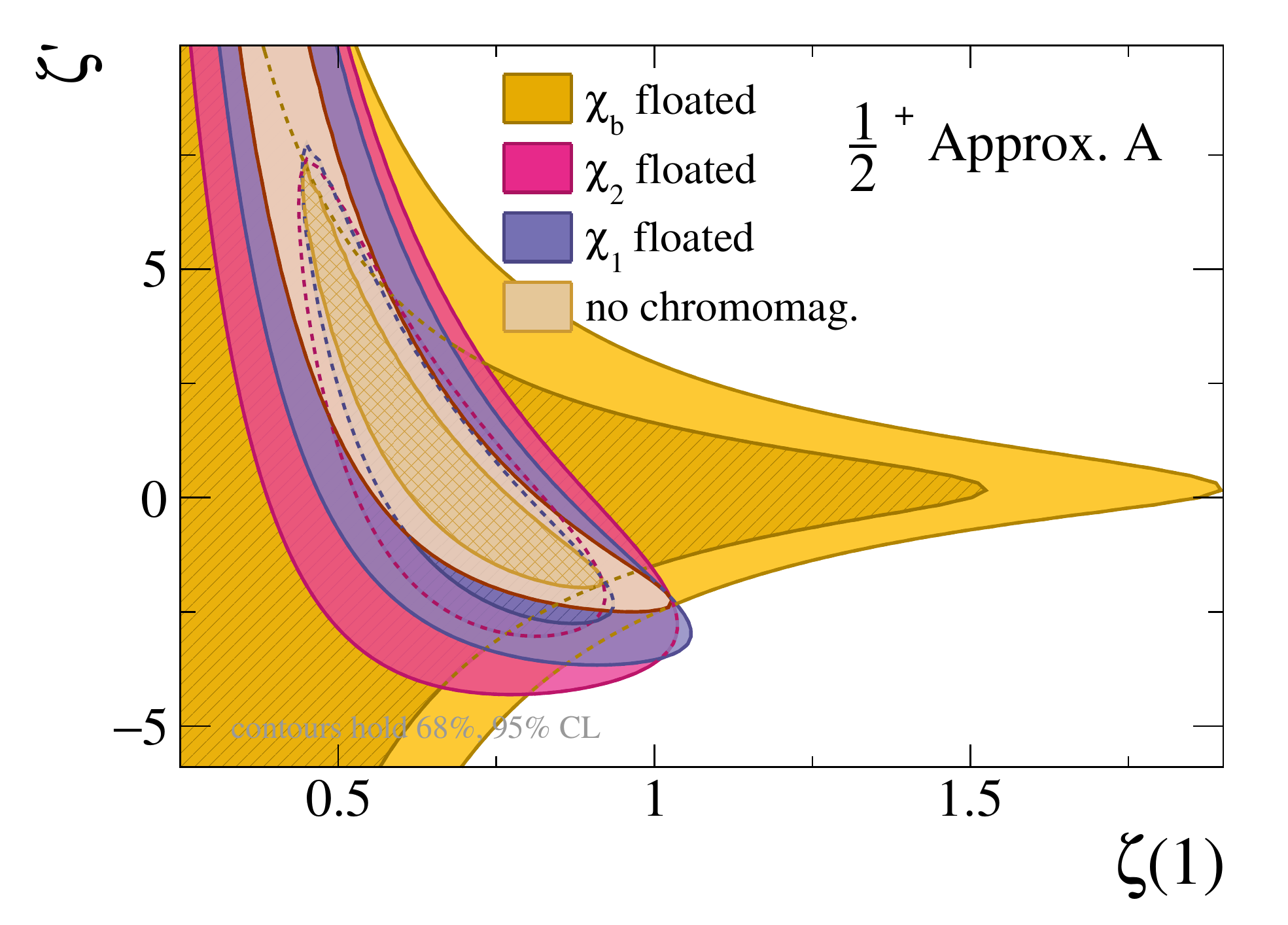} 
}
\centerline{
\includegraphics[width=0.5\textwidth]{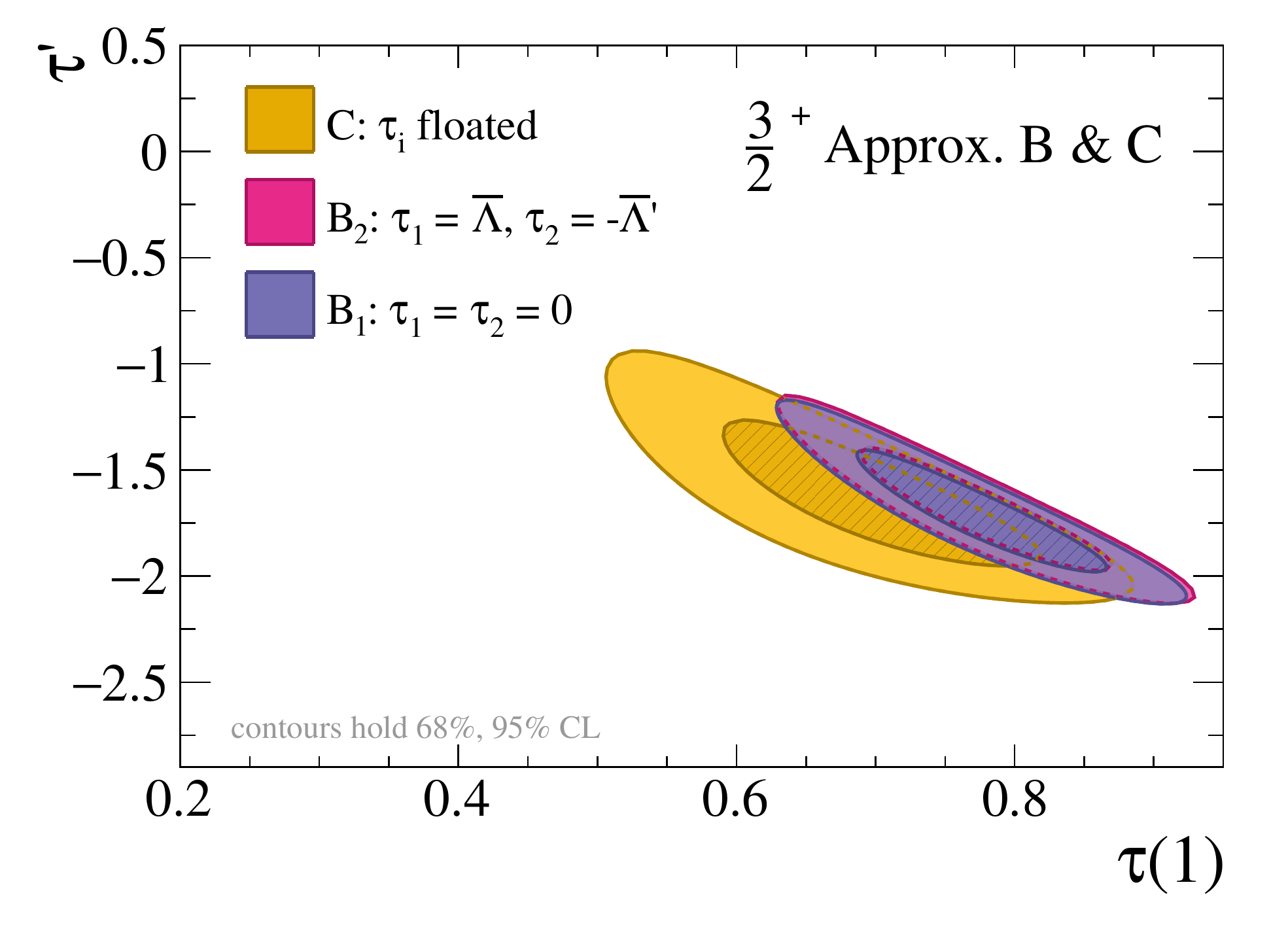}
\includegraphics[width=0.5\textwidth]{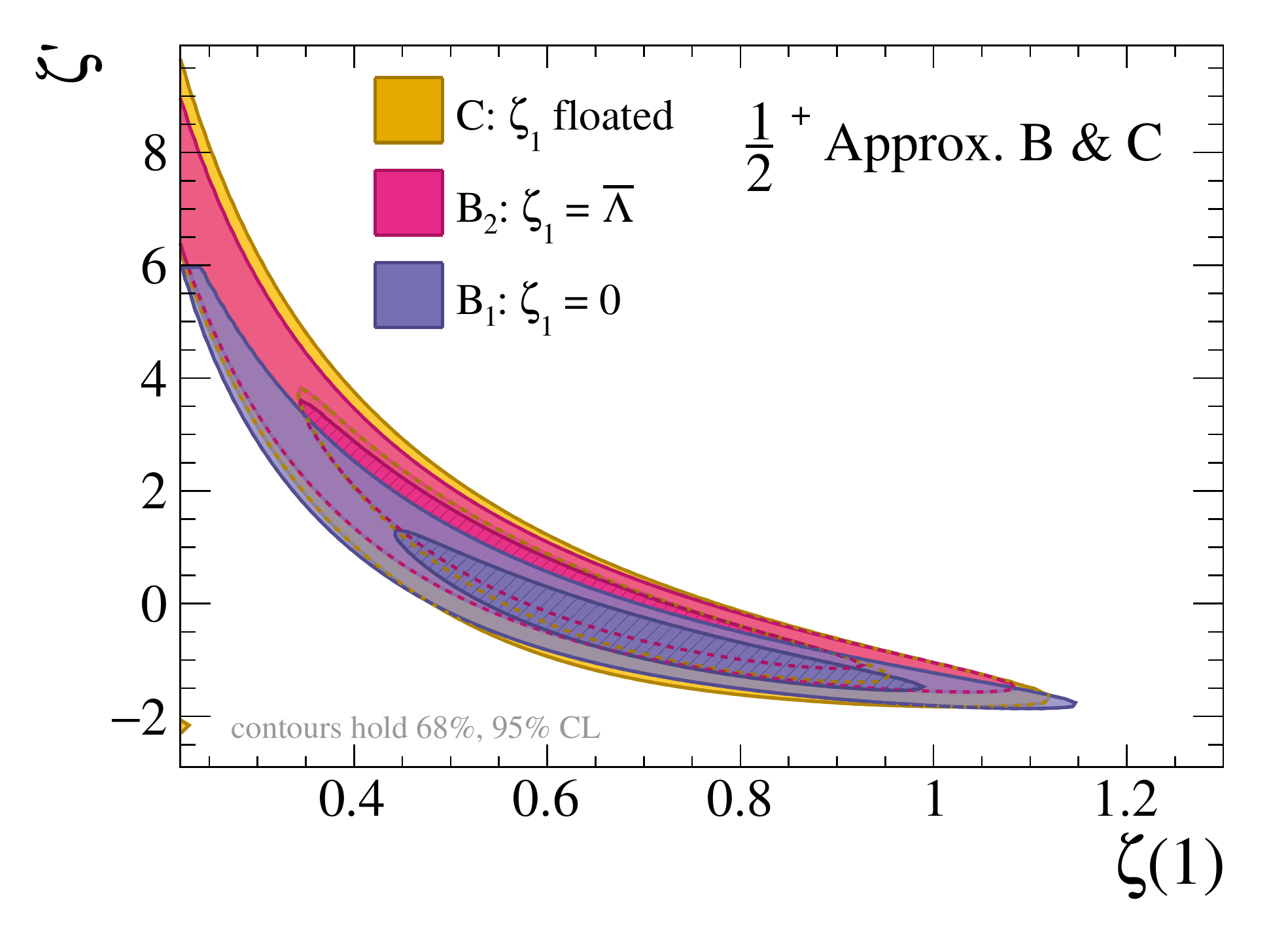}
}
\caption{The allowed 68\% and 95\% regions for $\tau(1)$ and $\tau'$ or
$\zeta(1)$ and $\zeta'$, respectively, are shown for the narrow $\frac{3}{2}^+$
(left) and broad $\frac{1}{2}^+$ states (right) for \ApproxA (top) and
\ApproxB (bottom).}
\label{fig:AppA_AppB_results}
\end{figure*}

Figure~\ref{fig:AppA_AppB_results} (top left) shows the 68\% and 95\% confidence
regions for the normalization and slope of the leading Isgur-Wise function for
the narrow $\frac32^+$ states. The scenarios explored are: no chromomagnetic
contributions, one chromomagnetic term (either $\eta_1$, $\eta_3$, or $\eta_b$;
note that $\eta_b$ and $\eta_1$ are degenerate in \ApproxA), or two
chromomagnetic terms (either $\eta_1$ or $\eta_b$ with $\eta_3$) marginalized.
Table~\ref{tab:AppA_results} summarizes the best fit points.  There is no
sensitivity to disentangle the different chromomagnetic contributions, and the
fitted values are compatible with zero. The extracted value for the slope of the
leading Isgur-Wise function is compatible with the $-1.5$ quark model prediction
in all scenarios.

Figure~\ref{fig:AppA_AppB_results} (top right) shows the 68\% and 95\%
confidence regions for the normalization and slope of the leading Isgur-Wise
function for the broad $\frac12^+$ states. The available experimental
information only loosely constrains the form factor parameters and introducing
one chromomagnetic contribution results only in marginal shifts of the extracted
normalization and slope of the leading Isgur-Wise function. The extracted value
for the slope of the leading Isgur-Wise function is compatible with quark model
predictions of $-1.0$ and the obtained chromomagnetic contributions are
compatible with zero. Table~\ref{tab:AppA_results} summarizes the extracted
best fit points. Table~\ref{tab:AppA_results2} summarizes the $\chi^2$ values of
all fits and the agreement of the best fit points with the experimental input is
good for the $\frac32^+$ states and $\frac12^+$ states for all scenarios.

\begin{table}[t!]
\begin{tabular}{l|ccc}
\hline\hline
  & $\tau(1)$ & $\tau'$ & $\eta_i$  \\
  \hline
 ---  & $0.63 \pm 0.02$  & $-1.29 \pm 0.17$ & ---   \\
 $\eta_1$  & $0.63  \pm 0.02$ & $ -1.53 \pm  0.52$ & $-0.10 \pm 0.19$   \\
 $\eta_3$  & $0.64 \pm 0.02$ & $-1.50 \pm 0.45$ &$ 0.14 \pm 0.29$  \\
 $\eta_b$  & $0.63  \pm 0.02$ & $ -1.53 \pm  0.52$ & $0.67 \pm 1.32$   \\
\hline\hline
\end{tabular}
\vspace{2ex}

\begin{tabular}{l|ccc}
\hline\hline
  & $\zeta(1)$ & $\zeta'$ & $\chi_i$  \\
  \hline
 ---  & $0.72 \pm 0.15$  & $-0.30 \pm 1.81$ & ---   \\
 $\chi_1$  & $0.73  \pm 0.15$ & $ -0.53 \pm  2.16$ & $0.03 \pm 0.15$   \\
 $\chi_2$  & $0.72 \pm 0.15$ & $-0.54 \pm 2.22$ &$ -0.05 \pm 0.30$  \\
\hline\hline
\end{tabular}
\caption{The best fit points of the \ApproxA fits, with and without
chromomagnetic contributions for the narrow $\frac{3}{2}^+$ (above) and broad
$\frac{1}{2}^+$ (below) states.}
\label{tab:AppA_results}
\end{table}

\begin{table}[t!]
\begin{tabular}{l | c cc}
\hline\hline
   & $\chi^2$ / ndf & Prob.  \\
  \hline
  ---  & 2.8 / 5 & 0.73  \\
 $\eta_1$ & 2.5 / 4 & 0.64   \\
 $\eta_3$  & 2.5 / 4 & 0.64  \\
 $\eta_b$ & 2.5 / 4 & 0.64  \\
\hline\hline
\end{tabular}
\hspace{2ex}
\begin{tabular}{l | c cc}
\hline\hline
   & $\chi^2$ / ndf & Prob.  \\
  \hline
  ---  & 8.7 / 6 & 0.19  \\
 $\chi_1$ & 8.7 / 5 & 0.12   \\
 $\chi_2$  & 8.7 / 5 & 0.12  \\
\hline\hline
\end{tabular}
\caption{The $\chi^2$ values and fit probabilities for the \ApproxA fits
for the narrow $\frac{3}{2}^+$ (left) and broad $\frac{1}{2}^+$ (right) states.}
\label{tab:AppA_results2}
\end{table}

Using the extracted values of the normalization and slope of the leading
Isgur-Wise function, and possible chromomagnetic contributions, the ratio of
semi-tauonic and semileptonic rates can be predicted. Including chromomagnetic
contributions change the central values of the predicted ratios only marginally,
but increase the uncertainties. Using the fitted values, we predict
\begin{align}
 R(D_2^*) = 0.06 \pm 0.01\,, & \qquad  \widetilde R(D_2^*) = 0.14 \pm 0.01\,,\nn\\*
 R(D_1) = 0.06 \pm 0.01\,,  & \qquad  \widetilde R(D_1) = 0.17 \pm 0.02\,, 
 \nn\\
 R(D_1^*) = 0.06 \pm 0.01\,,  & \qquad  \widetilde R(D_1^*) = 0.17 \pm 0.02\,,
 \nn\\
 R(D_0) =  0.07 \pm 0.03\,,  & \qquad  \widetilde R(D_0) = 0.22 \pm 0.04\,,
 \label{eq:AppA_RDds_results} 
\end{align}
and for the ratio of the sum of all four $D^{**}$ modes,
\begin{align}
 R(D^{**}) = 0.061 \pm 0.006 \,. \label{eq:AppA_RDds_results2}
\end{align}
The uncertainties are from the fit to the experimental information and also
contain the impact from possible chromomagnetic contributions.  In \ApproxA, the
predictions for $\widetilde R(D^{**})$ are more precise and more reliable than
for $R(D^{**})$, as the $w$ range is smaller. However, the experimental input to
make full use of this is not available yet, as partial branching fractions with a
$w$ cut would be needed.  Then the parameters in \ApproxA could be determined
just from the $q^2 > m_\tau^2$ part of phase space, resulting in better
precision for these predictions.

The obtained values can be compared to the prediction of the LLSW model. As
input we re-fit the normalization of the leading Isgur-Wise function
$\tau(1) = 0.64$ using the averaged semileptonic $D_1$ branching fraction from
Table~\ref{tab:sl_input}, and use
\begin{align}
 \zeta(1) = \frac{2}{\sqrt{3}}\, \tau(1)\,, \qquad \zeta' = \tau' + \frac12 \, ,
\end{align}
to relate the narrow $\frac{3}{2}^+$ and broad $\frac{1}{2}^+$ Isgur-Wise
functions. For the slope we use $\hat\tau' = -1.5$ discussed in
Section~\ref{sec:FF} based on model predictions. We find
\begin{align}
 R(D_2^*) = 0.06\,, & \qquad  \widetilde R(D_2^*) = 0.15\,,  \nn \\
 R(D_1) = 0.06\,,  & \qquad  \widetilde R(D_1) = 0.17\,,  \nn \\
 R(D_1^*) = 0.06\,,  & \qquad  \widetilde R(D_1^*) = 0.17\,, \nn \\
 R(D_0) =  0.08\,,  & \qquad  \widetilde R(D_0) = 0.23\,,
\end{align}
and for the ratio of the sum of all four $D^{**}$ modes,
\begin{align}
 R(D^{**}) = 0.064 \, ,
\end{align}
which are in excellent agreement with Eqs.~(\ref{eq:AppA_RDds_results}) and
(\ref{eq:AppA_RDds_results2}).

\subsection{\ApproxB and C}

The parameters of interest for \ApproxB are the normalization and slope
of the leading Isgur-Wise function, and the normalizations of the subleading
Isgur-Wise functions, $\tau_1$, $\tau_2$ or $\zeta_1$ (see
Section~\ref{sec:FF}). These parameters are again extracted separately for the
broad and narrow states using a simultaneous analysis of all semileptonic and
nonleptonic branching fractions. In addition, Approximations~B$_1$ and B$_2$
are explored, with the normalizations fixed. 

Figure~\ref{fig:AppA_AppB_results} (bottom left) shows the 68\% and 95\%
confidence regions for $\tau(1)$ and $\tau'$ for the narrow $\frac32^+$ states.
All three fit scenarios are in good agreement for the normalization and slope of
the leading Isgur-Wise function. Table~\ref{tab:AppB_results} summarizes the
best fit points and the obtained slope is compatible with the quark model
predictions used in Ref.~\cite{Leibovich:1997em}. Introducing the normalizations
of the subleading Isgur-Wise functions as free parameters, pulls them outside
the interval covered by Approximations~B$_1$ and B$_2$. This is interesting, as
in many experimental analyses the difference between Approximations~B$_1$ and
B$_2$ is used as a measure of the uncertainties associated with $D^{**}$
contributions. The \ApproxC parameter correlations for $\{ \tau(1), \tau', \tau_1, \tau_2 \}$. 
 are
\begin{align}
C_{\frac32^+} = \left( \begin{matrix}
    1 & - 0.83 & 0.66 & -0.63 \\
     - 0.83 & 1 & -0.27 & 0.20 \\
    0.66  & -0.27 & 1 & -0.93 \\
    -0.63  &  0.20 & -0.93 & 1
       \end{matrix} \right) \, . \, 
\end{align}

Figure~\ref{fig:AppA_AppB_results} (bottom right) shows the 68\% and 95\%
confidence regions for $\zeta(1)$ and $\zeta'$ for the broad $\frac12^+$ states.
There is good consistency of the normalizations and slopes of the leading
Isgur-Wise function, and the results for all three fits are listed in
Table~\ref{tab:AppB_results}. The normalization of the subleading Isgur-Wise
function, $\zeta_1$, is again outside the interval covered by
Approximations~B$_1$ and B$_2$. Table~\ref{tab:AppB_results2} summarizes the
compatibility of the best fit points, and the agreement is fair, with the
exception of the Approximation~B$_2$ fit for the narrow $\frac32^+$ states. The
\ApproxC parameter correlation for $\{ \zeta(1), \zeta', \zeta_1\}$ are
\begin{align}
C_{\frac12^+} = \left( \begin{matrix}
    1 & - 0.95 & -0.35 \\
    - 0.95   & 1  & 0.51 \\
    -0.35   & 0.51  &   1
       \end{matrix} \right) \, . \, 
\end{align}

\begin{table}[b!]
\begin{tabular}{l | cc cc}
\hline\hline
 & $\tau(1)$ & $\tau'$ & $\tau_1$ & $\tau_2$  \\
  \hline
 B$_1$  & $ 0.78 \pm 0.06$ & $-1.7 \pm 0.2$ & 0  & 0 \\
 B$_2$  & $ 0.78 \pm 0.06$ & $-1.7 \pm 0.2$  & $0.40$  & $-0.80$ \\
 C  & $ 0.71 \pm 0.07$ & $-1.6 \pm 0.2$  & $-0.5 \pm 0.3$  & $2.9 \pm 1.6$ \\
\hline\hline
\end{tabular}
\vspace{2ex}

\begin{tabular}{l | c cc}
\hline\hline
   & $\zeta(1)$ & $\zeta'$ & $\zeta_1$  \\
  \hline
 B$_1$  & $0.73 \pm 0.18$ & $-0.7 \pm 0.8$ & 0 \\
 B$_2$  & $0.66 \pm 0.19$ & $ 0 \pm 1.1$ & $0.4$ \\
 C  & $0.68 \pm 0.20$ & $-0.2 \pm 1.2$ & $0.3 \pm 0.3$ \\
\hline\hline
\end{tabular}
\caption{The best fit points of the \ApproxB and C fits for the narrow
$\frac{3}{2}^+$ (above) and broad $\frac{1}{2}^+$ (below) states.}
\label{tab:AppB_results}
\end{table}

\begin{table}[b!]
\begin{tabular}{l | c cc}
\hline\hline
   & $\chi^2$ / ndf & Prob.  \\
  \hline
  B$_1$  & $6.1 /  6$ & 0.42  \\
  B$_2$  & $11.6 /  6$ & 0.07 \\
  C  & $2.4 / 4 $ & 0.66 \\
\hline\hline
\end{tabular}
\hspace{2ex}
\begin{tabular}{l | c cc}
\hline\hline
   & $\chi^2$ / ndf & Prob.  \\
  \hline
  B$_1$  & $10.1/ 5$ & 0.07 \\
  B$_2$  & $9.2/ 5 $ & 0.10 \\
  C   & $9.1/ 4 $ & 0.06 \\
\hline\hline
\end{tabular}
\caption{The $\chi^2$ values and fit probabilities for the \ApproxB and C
fits for the narrow $\frac{3}{2}^+$ (left) and broad $\frac{1}{2}^+$ states
(right).}
\label{tab:AppB_results2}
\end{table}

\begin{figure*}[th]
\centerline{
\includegraphics[width=0.48\textwidth]{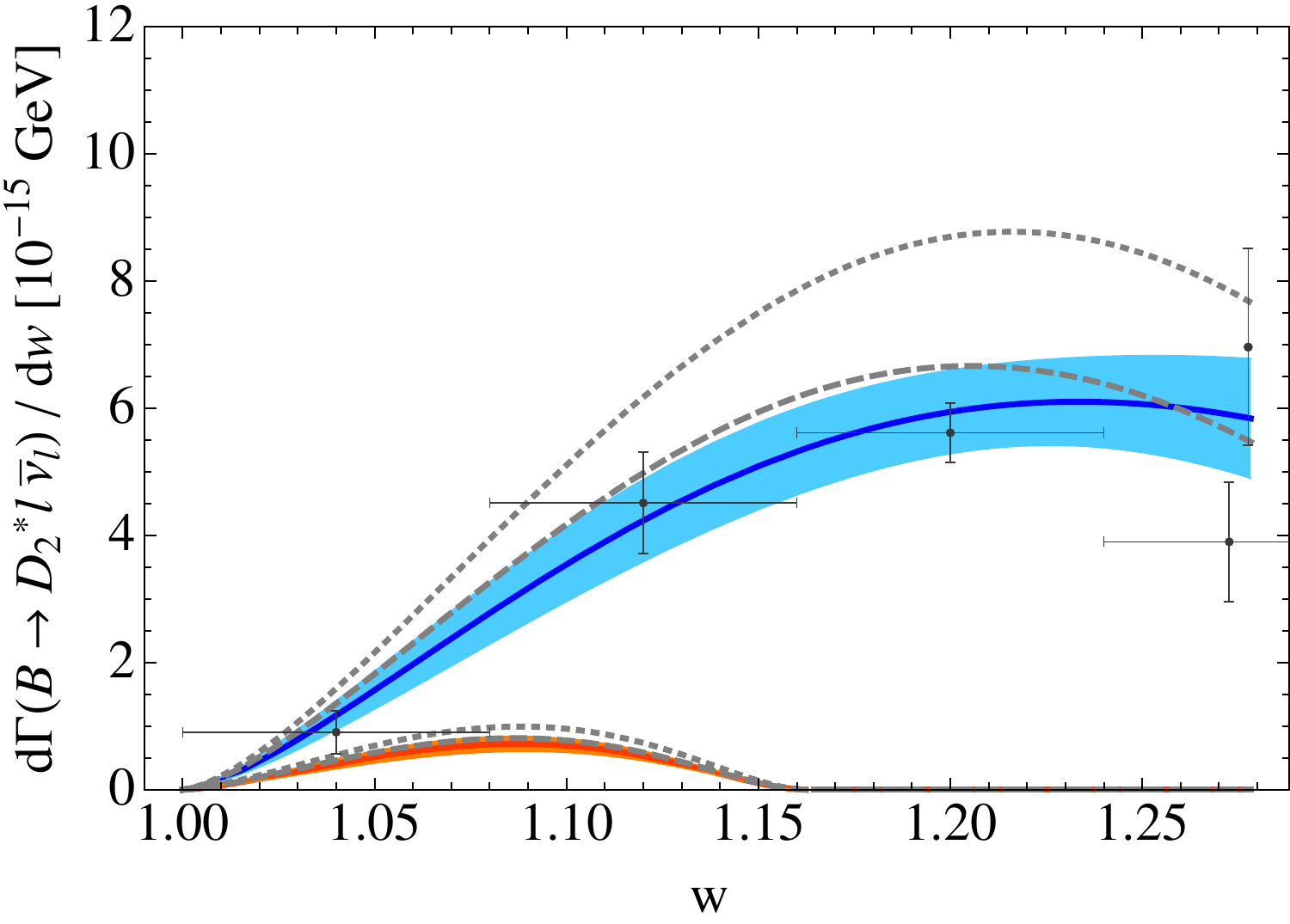}\hfill
\includegraphics[width=0.48\textwidth]{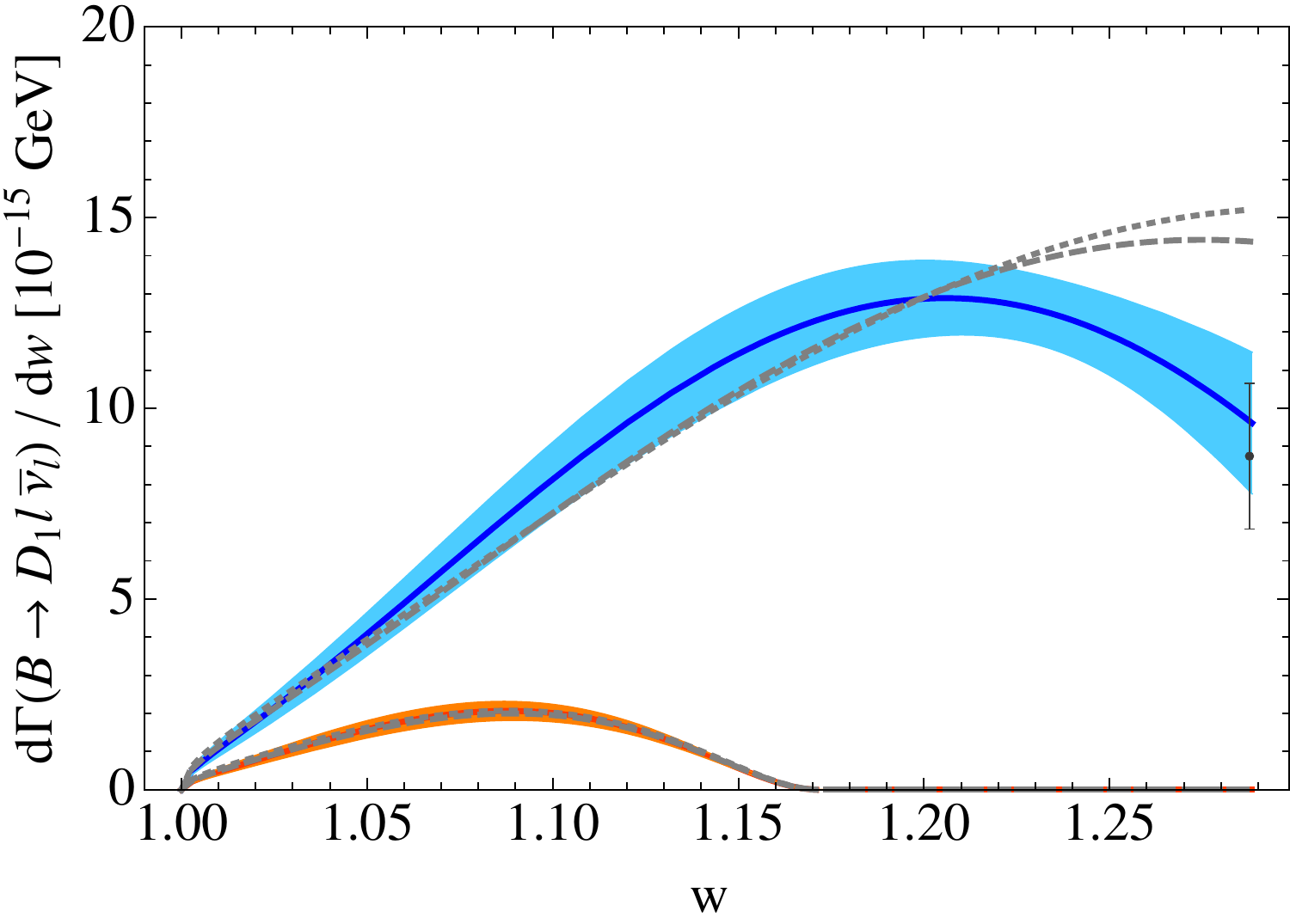} 
}
\centerline{
\includegraphics[width=0.48\textwidth]{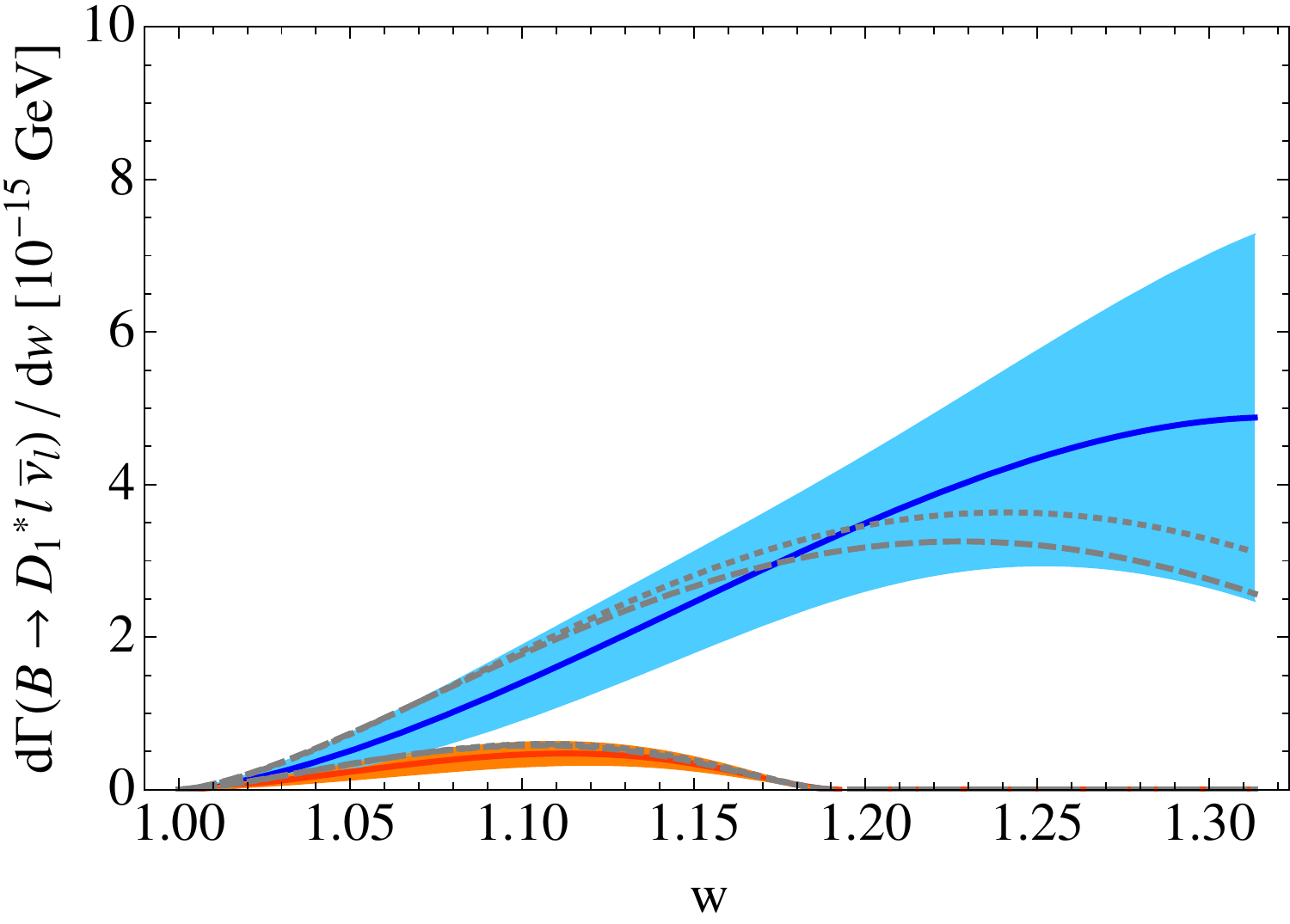}\hfill
\includegraphics[width=0.48\textwidth]{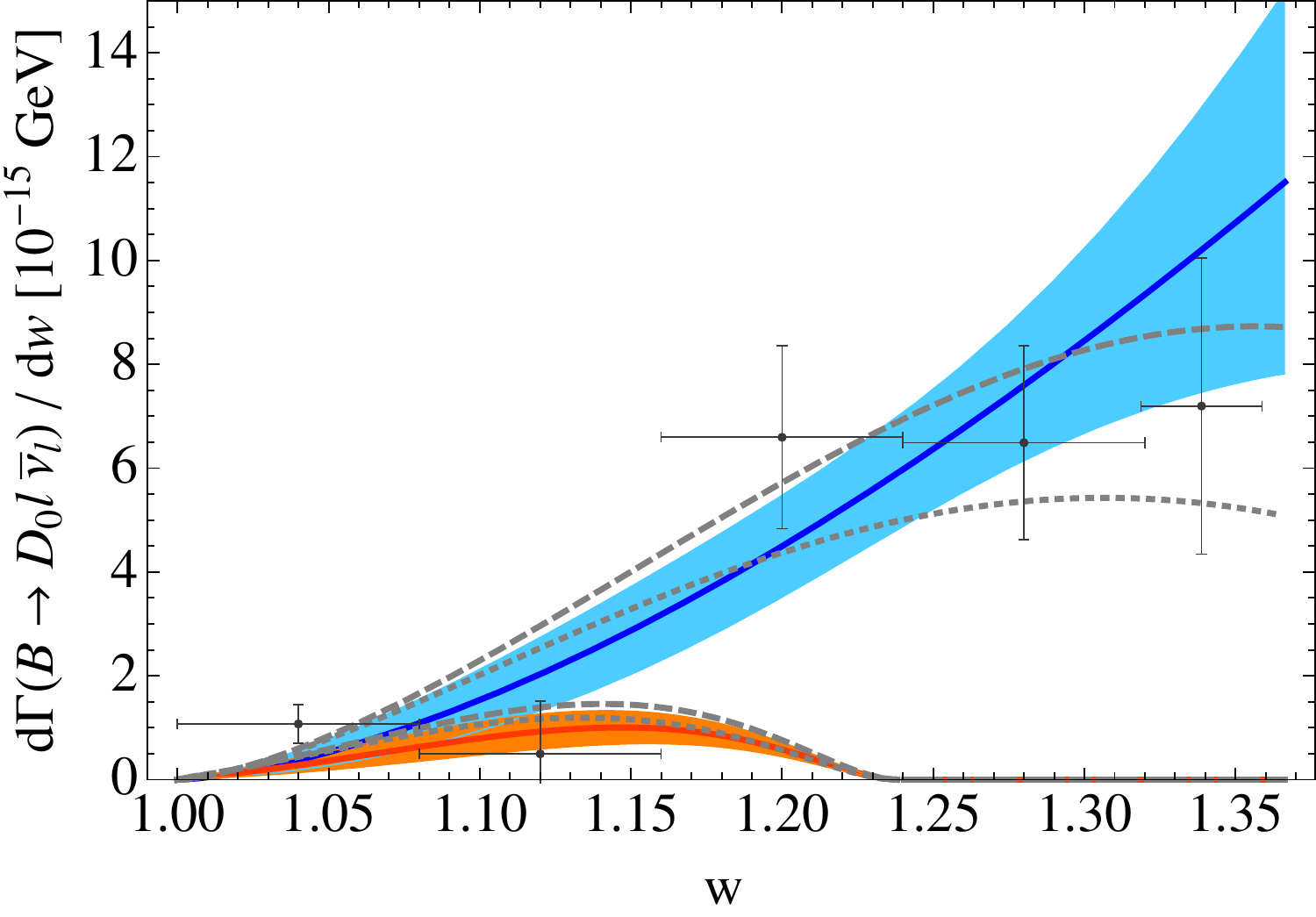}
}
\caption{The colored bands show the allowed 68\% regions for $m_\ell = 0$ (blue)
and $m_\ell = m_\tau$ (orange) for the differential decay rates in \ApproxC. The
dashed (dotted) curves show the predictions of Ref.~\cite{Leibovich:1997em} for
Approximations~B$_1$ (B$_2$). The data points correspond to the differential
semileptonic or nonleptonic branching fraction measurements described in the
text.}
\label{fig:AppB_rates}
\end{figure*}

Using the fit results, with the normalizations of the subleading Isgur-Wise
functions floated, in \ApproxC we obtain
\begin{align}\label{eq:AppB_RDds_results}
 R(D_2^*) = 0.07 \pm 0.01\,,  & \qquad  \widetilde R(D_2^*) = 0.17\pm 0.01\,, \nn  \\*
 R(D_1) = 0.10 \pm 0.01\,,  & \qquad  \widetilde R(D_1) = 0.20 \pm 0.01\,, \nn \\*
 R(D_1^*) = 0.06 \pm 0.02\,,  & \qquad  \widetilde R(D_1^*) = 0.18 \pm 0.02\,, \nn \\*
 R(D_0) = 0.08 \pm 0.03\,,  & \qquad  \widetilde R(D_0) = 0.25 \pm 0.03\,,
\end{align}
and for the ratio for the sum over all four $D^{**}$ states,
\begin{equation}
R(D^{**}) = 0.085\pm 0.010 \,. \label{eq:AppB_RDds_results2}
\end{equation}
These values can be compared with the LLSW prediction, including the lepton mass
effects in Eqs.~(\ref{D1rate}), (\ref{D2rate}), and (\ref{D0rate}). Using
Eq.~(\ref{eq:leading_IW}) for the Isgur-Wise functions for the $\frac{3}{2}^+$
states, and the model prediction in Eq.~(\ref{eq:quark_model_IW}) to relate it
to the $\frac{1}{2}^+$ states, we find in Approximation~B$_1$ and B$_2$, respectively,
\begin{align}
 R(D_2^*) = \{0.072,\, 0.068\},  &\qquad  \widetilde R(D_2^*) = \{0.159,\, 0.158\},  \nn \\
 R(D_1) = \{0.096,\, 0.099\},  &\qquad  \widetilde R(D_1) = \{0.221,\, 0.231\}, \nn \\
 R(D_1^*) = \{0.092,\, 0.083\},  & \qquad  \widetilde R(D_1^*) = \{0.200,\, 0.196\}, \nn \\
 R(D_0) = \{0.107,\, 0.118\}, & \qquad  \widetilde R(D_0) = \{0.272,\, 0.275\},
\end{align}
and for the sum of the four $D^{**}$ states,
\begin{equation}
 R(D^{**}) =  \{0.0949,\, 0.0946\}\,.
\end{equation}
The ranges spanned by these Approximation~B$_1$ and B$_2$ results do not
necessarily give conservative estimates of the uncertainties. These ratios,
however, are in good agreement with Eqs.~(\ref{eq:AppB_RDds_results}) and
(\ref{eq:AppB_RDds_results2}).

Of the mass parameters, $\bar\Lambda$ has substantially bigger uncertainty than
$\bar\Lambda' - \bar\Lambda$ or $\bar\Lambda^* - \bar\Lambda$.  Varying
$\bar\Lambda$ by $\pm50\,\MeV$ while keeping the differences fixed has a
negligible impact compared to other uncertainties included.  This is consistent
with the fact that in \ApproxA the only dependence on $\bar\Lambda$ is via
$\bar\Lambda' - \bar\Lambda$ and $\bar\Lambda^* - \bar\Lambda$.

Figure~\ref{fig:AppB_rates} shows the differential decay rates of the \ApproxC
fits as functions of $w$ for $m_\ell = 0$ and $m_\ell = m_\tau$, with the
corresponding 68\% uncertainty bands. The LLSW model
prediction is also shown for the differential decay rates: the dashed (dotted)
curves show Approximation~B$_1$ (B$_2$) and the normalization of the
leading Isgur-Wise function was determined using the averaged semileptonic
$D_1$ branching fraction, which gives $\tau(1) = 0.80$. The \ApproxC fit using
the full differential semileptonic and nonleptonic information constrain the
shape stronger than the LLSW model, which only uses the $D_1$ rate information.

\begin{figure*}[th]
\centerline{
\includegraphics[width=0.48\textwidth]{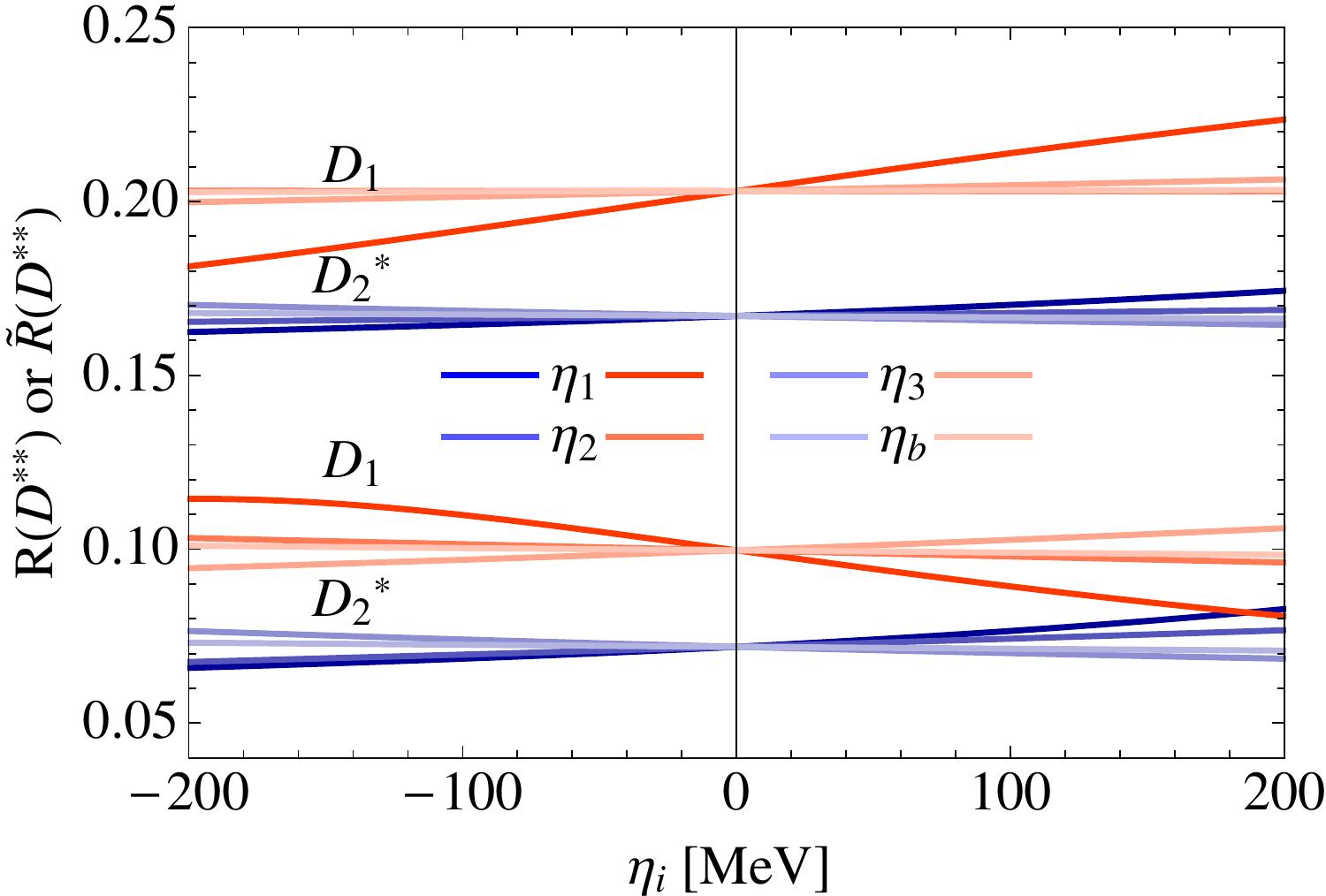}\hfill
\includegraphics[width=0.48\textwidth]{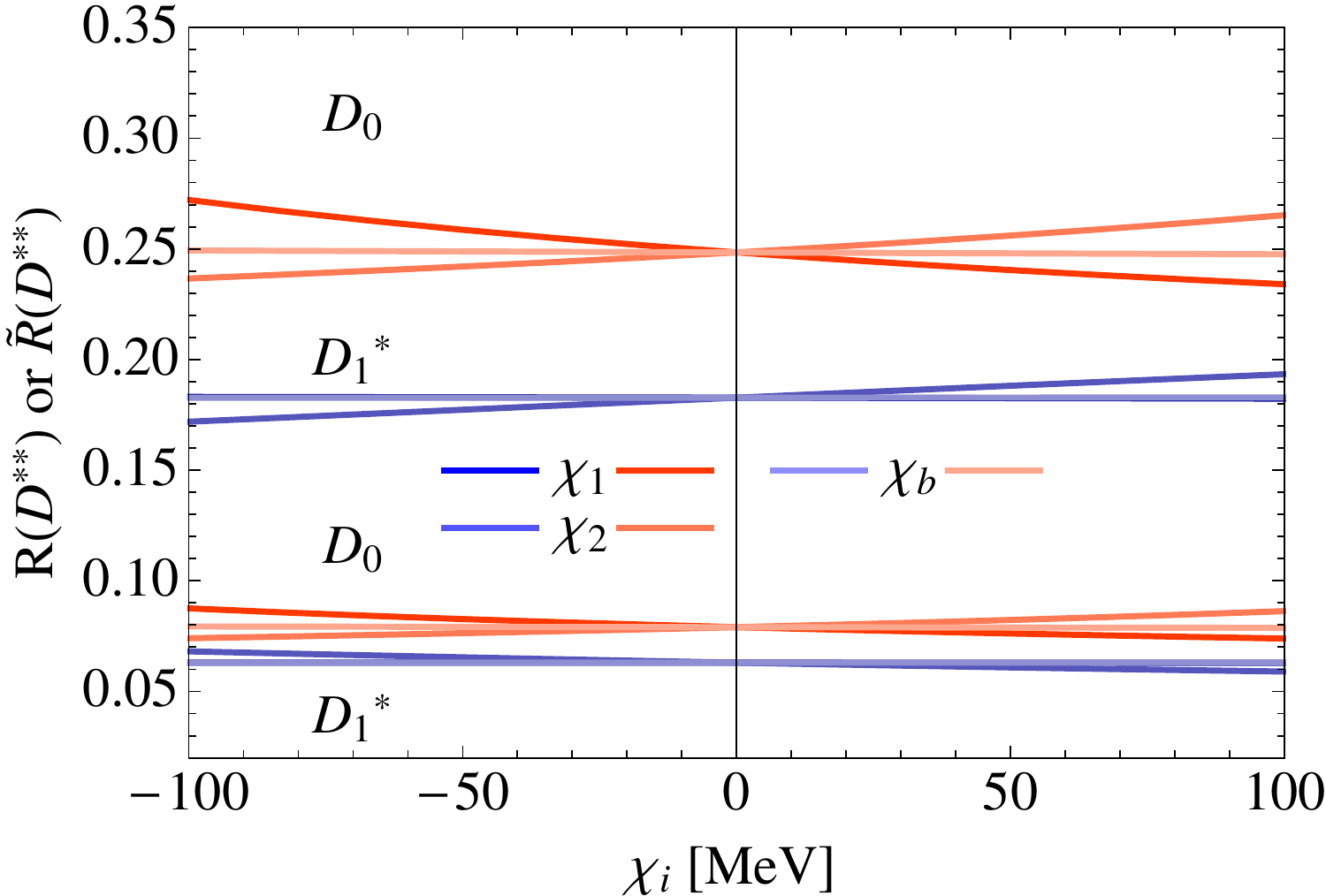}
}
\caption{The impact of chromomagnetic contributions $\eta_i$ and $\chi_i$ on the
exclusive ratios $R(D^{**})$ (below 0.15) and  $\widetilde R(D^{**})$ (above
0.15). For the leading and subleading Isgur-Wise functions the best fit
parameters of \ApproxC (without including chromomagnetic terms) are used. The
explored range is motivated by the experimental constraints of $\eta_1$ and
$\chi_1$ (see the text).}
\label{fig:App_C_Chromo}
\end{figure*}

We also explore in \ApproxC the impact of additional chromomagnetic
contributions. The available experimental information does not allow to
disentangle subleading Isgur-Wise function contributions from chromomagnetic
terms. Figure~\ref{fig:App_C_Chromo} shows the dependence of $R(D^{**})$ on
one of the chromomagnetic contributions at a time.  For the narrow
$\frac32^+$ states the only strong dependence comes from $\eta_1$. This
originates from large factors in the rate expressions, and if introduced as
an additional free parameter in the \ApproxC fit, its size is constrained to be
about $\pm 200$ MeV, but it is also strongly correlated to other subleading
Isgur-Wise function normalizations. For the broad $\frac12^+$ states the
strongest dependence comes from $\chi_1$. If introduced as an additional free
parameter in the \ApproxC fit, its size is constrained to be about $\pm 100$
MeV.

To account for these subleading Isgur-Wise functions parameterizing
chromoagnetic effects, we can recalculate the ratios of semi-tauonic and
semileptonic rates by introducing an additional uncertainty of $\pm 200$ MeV
and  $\pm 100$ MeV on $\eta_1$ and $\chi_1$, respectively. We thus obtain
\begin{align}\label{eq:AppC_RDds_results}
 R(D_2^*) = 0.07 \pm 0.01\,,  & \qquad  \widetilde R(D_2^*) = 0.17\pm 0.01\,, \nn  \\
 R(D_1) = 0.10 \pm 0.02\,,  & \qquad  \widetilde R(D_1) = 0.20 \pm 0.02\,,  \nn \\
 R(D_1^*) = 0.06 \pm 0.02\,,  & \qquad  \widetilde R(D_1^*) = 0.18 \pm 0.02\,, \nn \\
 R(D_0) = 0.08 \pm 0.04\,, & \qquad  \widetilde R(D_0) = 0.25 \pm 0.06\,,
\end{align}
and for the ratio of the sum over all four $D^{**}$ states,
\begin{align}
R(D^{**}) = 0.085\pm 0.012 \, . \label{eq:AppC_RDds_results2}
\end{align}
These uncertainties are not much greater than those in
Eqs.~(\ref{eq:AppB_RDds_results}) and (\ref{eq:AppB_RDds_results2}).

\section{$B_s \to D_s^{**} \ell \, \bar\nu$ decays}
\label{sec:SMBs_rate}

An important difference between $B\to D^{**}\ell\bar\nu$ and $B_s\to
D_s^{**}\ell\bar\nu$ is that the two lightest excited $D_s$ states observed are
fairly narrow.  They are lighter than the $m_{D^{(*)}} + m_K$ mass thresholds,
so they can only decay to $D_s^{(*)}\pi$, which violate isospin (if these are
the $D_s^{**}$ isosinglet $s_l^{\pi_l} = \frac12^+$ orbitally excited states). 
Due to these narrow widths, semi-tauonic $B_s$ decay to the spin-zero meson,
$B_s\to D_{s0}^*\tau\bar\nu$, may be easier to measure than $B\to
D_0^*\tau\bar\nu$, and may provide good sensitivity to possible scalar
interactions from new physics.\footnote{We thank Marcello Rotondo for drawing
our attention to this.} Table~\ref{tab:charmstrange} summarizes the relevant
masses  and widths. 

While the $s_l^{\pi_l} = \frac32^+$ doublets in both the $D_s^{**}$ and
$B_s^{**}$ cases have masses ``as expected", about 100\,MeV above their
non-strange counterparts, the masses of the $s_l^{\pi_l} = \frac12^+$ doublet of
$D_s^{**}$ states are surprisingly close to their non-strange counterparts. 
(Which is why the discovery of the $D_{s0}^*$~\cite{Aubert:2003fg} was such a
surprise.)  This unexpected spectrum makes the analysis in this Section more
uncertain than in the previous ones.  

It is possible that interpreting the $D_{s0}^*$ and $D_{s1}^*$ as the lightest
orbitally excited states is oversimplified (and this is what our description
assumes), and we have higher confidence that our description of the decays to
the $s_l^{\pi_l} = \frac32^+$ $D_{s1}$ and $D_{s2}^*$ states should be
reliable.  The first exploratory lattice QCD studies that obtain the $D_{s0}^*$
and $D_{s1}^*$ masses in agreement with data appeared only
recently~\cite{Leskovec:2015naf}.  To be more specific, assuming that the
$D_{s0}^*$ is the lightest orbitally excited $D_s$ state, theoretical
predictions for ${\cal B}(D_{s0}\to D_s^*\gamma) / {\cal B}(D_{s0}\to D_s\pi)$
tend to be above~\cite{Godfrey:2003kg, Colangelo:2003vg, Mehen:2004uj} the CLEO
upper bound, ${\cal B}(D_{s0}\to D_s^*\gamma) / {\cal B}(D_{s0}\to D_s\pi) <
0.059$ (90\%~CL)~\cite{Besson:2003cp}.  The $D^{(*)}K$ molecular picture of
these states also faces challenges, e.g., the lack of observed isospin
partners~\cite{Choi:2015lpc}.  It is possible that the correct description is a
mixture of these.  However, given that the CLEO bound~\cite{Besson:2003cp} was
obtained with 13.5/fb data, and the Belle bound on the above ratio $<0.18$
(90\%~CL)~\cite{Abe:2003jk} used 87/fb, while the \babar result $<0.16$
(95\%~CL)~\cite{Aubert:2006bk} used 232/fb, remeasuring  ${\cal B}(D_{s0}\to
D_s^*\gamma) / {\cal B}(D_{s0}\to D_s\pi)$ using the full \babar and Belle data
would be desirable.

Another piece of data is that the mass splittings within each heavy quark spin
symmetry doublets appear to be consistent with nominal $SU(3)$ breaking between
the strange and non-strange states.  This supports the fact that the mass
splittings in the $s_l^{\pi_l} = \frac12^+$ doublets are comparable to $m_{D^*}
- m_D \simeq m_{D_s^*} - m_{D_s}$, unlike what LLSW considered based on the data
in 1997.

\begin{table}[bt]
\begin{tabular}{ccccc}
\hline\hline
Particle  &  $s_l^{\pi_l}$ &  $J^P$  &  $m$ (MeV)  &  $\Gamma$ (MeV)\\
\hline
$D_{s0}^*$ &  $\frac12^+$  &  $0^+$  &  $2318$  &  $<4$ \\
$D_{s1}^*$ &  $\frac12^+$  &  $1^+$  &  $2460$  &  $<4$ \\[2pt]
\hline
$D_{s1}$ &  $\frac32^+$  &  $1^+$  &  $2535$  &  1 \\
$D_{s2}^*$ &  $\frac32^+$  &  $2^+$  &  $2567$  &  17 \\[2pt]
\hline\hline
$B_{s1}$ &  $\frac32^+$  &  $1^+$  &  $5829$  &  1 \\
$B_{s2}^*$ &  $\frac32^+$  &  $2^+$  &  $5840$  &  1 \\[2pt]
\hline\hline
\end{tabular}
\caption{Same as Table~\ref{tab:charm}, but for $D_s$ mesons.
For the $\frac32^+$ states we averaged the PDG with a recent LHCb
measurement~\cite{Aaij:2016utb} not included in the PDG.}
\label{tab:charmstrange}
\end{table}

\begin{table}[tb]
\begin{tabular}{ccc|cc}
\hline\hline
$s_l^{\pi_l}$ & Particles & $\overline{m}$ (MeV) & Particles & $\overline{m}$ (MeV) \\ \hline 
$\frac12^-$  &  $D_s$, $D_s^*$	&  2076 & $B_s$, $B_s^*$  &  5403  \\
$\frac12^+$  &  $D_{s0}^*$, $D_{s1}^*$ & 2425 &  $B_{s0}^*$, $B_{s1}^*$ & ---\\
$\frac32^+$  &  $D_{s1}$, $D_{s2}^*$ & 2555  &  $B_{s1}$, $B_{s2}^*$  & 5836 \\
\hline\hline
\end{tabular}
\caption{Same as Table~\ref{tab:averagemass}, but for $D_s$ and $B_s$ mesons.} 
\label{tab:averagemass2}
\end{table}

For the HQET mass parameters we use $\bar\Lambda_s = \bar\Lambda + 90$\,MeV,
motivated by $\ov m_{B_s} - \ov m_B$.  We also estimate $\bar\Lambda_s' -
\bar\Lambda_s = 0.41$\,GeV using Eq.~(1.10) in Ref.~\cite{Leibovich:1997em}. 
For $\bar\Lambda_s' - \bar\Lambda_s^*$ we estimate $0.13$\,GeV from the $(2555-2425)$\,MeV
difference in Table~\ref{tab:averagemass2}.  (These values are also shown in
Table~\ref{tab:input_summary}.). 

Using $SU(3)$ flavor symmetry to relate the $B\to D^{**}\ell\bar\nu$ decay
parameters to $B_s\to D_s^{**}\ell\bar\nu$, in \ApproxC we predict for the
ratios of the $\tau$ to light-lepton rates
\begin{align}\label{eq:AppC_RDsds_results}
 R(D_{s2}^*) = 0.07 \pm 0.01\,,  & \qquad  \widetilde R(D_{s2}^*) = 0.16\pm 0.01\,, \nn  \\*
 R(D_{s1}) = 0.09 \pm 0.02\,,  & \qquad  \widetilde R(D_{s1}) = 0.20 \pm 0.02\,,  \nn \\*
 R(D_{s1}^*) = 0.07 \pm 0.03\,,  & \qquad  \widetilde R(D_{s1}^*) = 0.20 \pm 0.02\,, \nn \\*
 R(D_{s0}^*) = 0.09 \pm 0.04\,, & \qquad  \widetilde R(D_{s0}^*) = 0.26 \pm 0.05\,.
\end{align}
This is the analog of Eq.~(\ref{eq:AppC_RDds_results}), with increased
uncertainties to account for the impact of additional chromomagnetic
contributions, which cannot be constrained well yet.  These predictions will
improve when more data is available on $B\to D^{**}\ell\bar\nu$, or
$B_s\to D_s^{**}\ell\bar\nu$, or related $B_{(s)}\to D_{(s)}^{**}\pi$ rates.

\section{$B \to D^{**} \tau\, \bar\nu$ and scalar interactions}
\label{sec:2HDM}

To illustrate the complementary sensitivities to new physics, in this section we
explore the impacts of possible scalar interactions on $R(D^{**})$.  We consider
the effective Hamiltonian,
\begin{align}\label{Hnp}
{\cal H} = \frac{4G_F}{\sqrt2}\, & V_{cb}\, \big[
  (\bar c \gamma_\mu P_L b)\,(\bar\tau \gamma^\mu P_L \nu) \\
&\quad + S_L (\bar c P_L b)\,(\bar\tau P_L \nu)
  + S_R (\bar c P_R b)\,(\bar\tau P_L \nu) \big] . \nn
\end{align}
This notation follows Ref.~\cite{Lees:2013uzd}, although without specifying
details of the underlying model, one may expect $S_{L,R} = {\cal O} [m_W^2 /
(|V_{cb}|\Lambda^2)]$.  For simplicity, we consider two scenarios. (i) In the
type-II 2HDM, $S_L = 0$ and $S_R = - m_b\, m_\tau\tan^2\beta/m_{H^\pm}^2$.  This
scenario is motivated by being the Higgs sector of the MSSM; it does not give a
good fit to the current data, however, the central values of $R(D^{(*)})$ may
change in a way that  this conclusion is altered, but a robust deviation from
the SM remains.  (ii)~In another scenario that we consider, $S_L + S_R = 0.25$
and we study the dependence of $R(D^{**})$ on $S_L - S_R$.  This is motivated by
giving a good fit the current data, and can arise, e.g., in other extensions of
the Higgs sector.

While Eq.~(\ref{Hnp}) is natural to write in terms of left- and right-handed
operators, the hadronic matrix elements are simpler to address in terms of the
scalar ($\bar c b$) and pseudoscalar ($\bar c \gamma_5 b$) currents.  In
particular,
\begin{align}
\langle D_0^*| \bar c b |B \rangle &= 0\,, \qquad
\langle D_1^*| \bar c \gamma_5 b |B \rangle = 0\,, \nn\\*
\langle D_2^*| \bar c b |B \rangle &= 0\,, \qquad
\langle D_1| \bar c \gamma_5 b |B \rangle = 0\,,
\end{align}
whereas $\langle D^*| \bar c b |B \rangle = 0$ and
$\langle D| \bar c \gamma_5 b |B \rangle = 0$ for the $D$ and $D^*$.
The non-vanishing (pseudo)scalar matrix elements can be related to those of
the SM currents via
\begin{align}
\langle X| \bar c \gamma_5 b |B \rangle &= \frac{-q^\mu}{m_b+m_c}\, 
  \langle X| \bar c \gamma_\mu \gamma_5 b |B \rangle\,, \nn\\*
\langle X| \bar c b |B \rangle &= \frac{q^\mu}{m_b-m_c}\, 
  \langle X| \bar c \gamma_\mu b |B \rangle\,.
\end{align}

The charged Higgs contribution is simplest to include by writing
the rate in terms of a helicity decomposition.  The differential decay rate
with its full lepton mass dependence can be written as
\begin{align}\label{eq:rate}
\frac{\text{d} \Gamma(B \to D^{**} \ell\, \bar\nu)}{ \text{d} q^2}
  = \frac{G_F^2\, |V_{cb}|^2\, |\vec p\,'|\, q^2}{96 \pi^3\, m_B^2} 
  \bigg( 1 - \frac{m_\ell^2}{q^2} \bigg)^2 \nn \\
  \quad \times \bigg[ \sum_{k = \pm,0,t} H_k^2\,
  \bigg( 1 + \frac{m_\ell^2}{2 q^2} \bigg) + \frac32 \frac{m_\ell^2}{q^2} H_t^2
  \bigg] \,,
\end{align}
with the helicity amplitudes $H_{k = \pm,0,t}$ (we use the notation of
Ref.~\cite{Korner:1989qb}).  Here $|\vec p\,'|$ is the magnitude
of the three-momentum of the $D^{**}$. It is related to $q^2$ as
\beq
|\vec p\,'|^2 = \bigg( \frac{m_B^2 + m_{D^{**}}^2 - q^2}{2 m_B} \bigg)^2
  - m_{D^{**}}^2 = m_{D^{**}}^2 (w^2-1) \,.
\eeq
Setting $m_\ell=0$, one recovers the expression 
\beq\label{eq:rate_nomass}
\frac{\text{d} \Gamma(B \to D^{**} \ell \bar\nu)}{ \text{d} q^2} = 
  \frac{G_F^2\, |V_{cb}|^2\, |\vec p\,'|\, q^2}{96\pi^3\, m_B^2}
  \sum_{k=\pm,0,t} \! H_k^2 \,,
\eeq
which is an excellent approximation for $l =e,\mu$. 

The contributions of the scalar operators can be included by replacing
$H_t$ according to
\beq \label{eq:2hdmtypeIII}
H_t \to H_t^{\rm SM} \bigg[ 1 + (S_R \pm S_L)\,
  \frac{q^2}{m_\tau(m_b \mp m_c)} \bigg]\,,
\eeq
where the upper signs are for the final states $D$, $D_1^*$ and $D_1$, and the
lower signs are for $D^*$, $D_0^*$, and $D_2^*$.
The helicity amplitudes $H_{\pm,0,t}$ are related to the form factors defined in
Eqs.~(\ref{formf32}) and (\ref{formf12}), and the full expressions for all
four $D^{**}$ states are given in Appendix~\ref{App:helamps}.

\begin{figure}[t]
\includegraphics[width=\columnwidth]{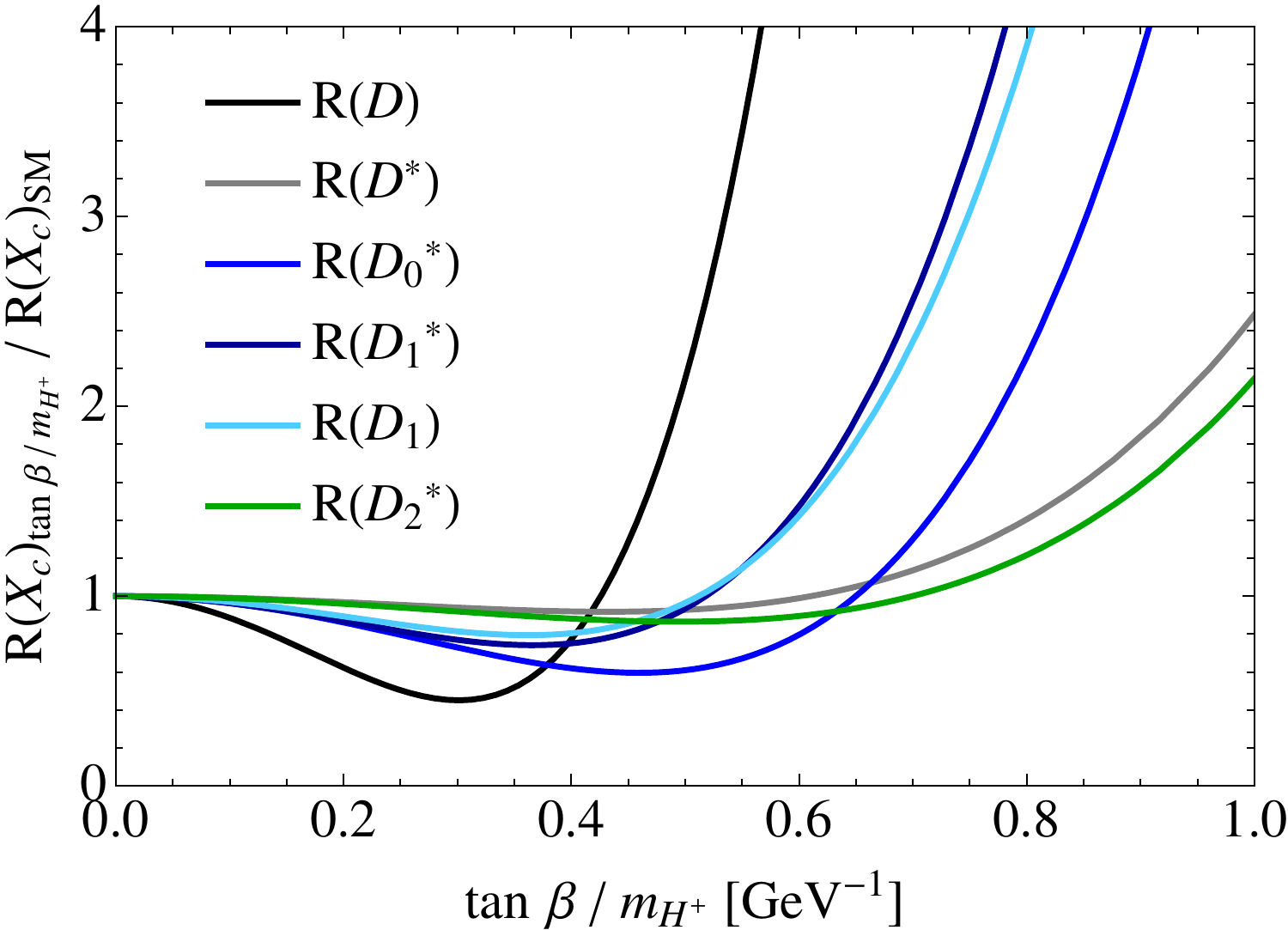}
\includegraphics[width=0.97\columnwidth]{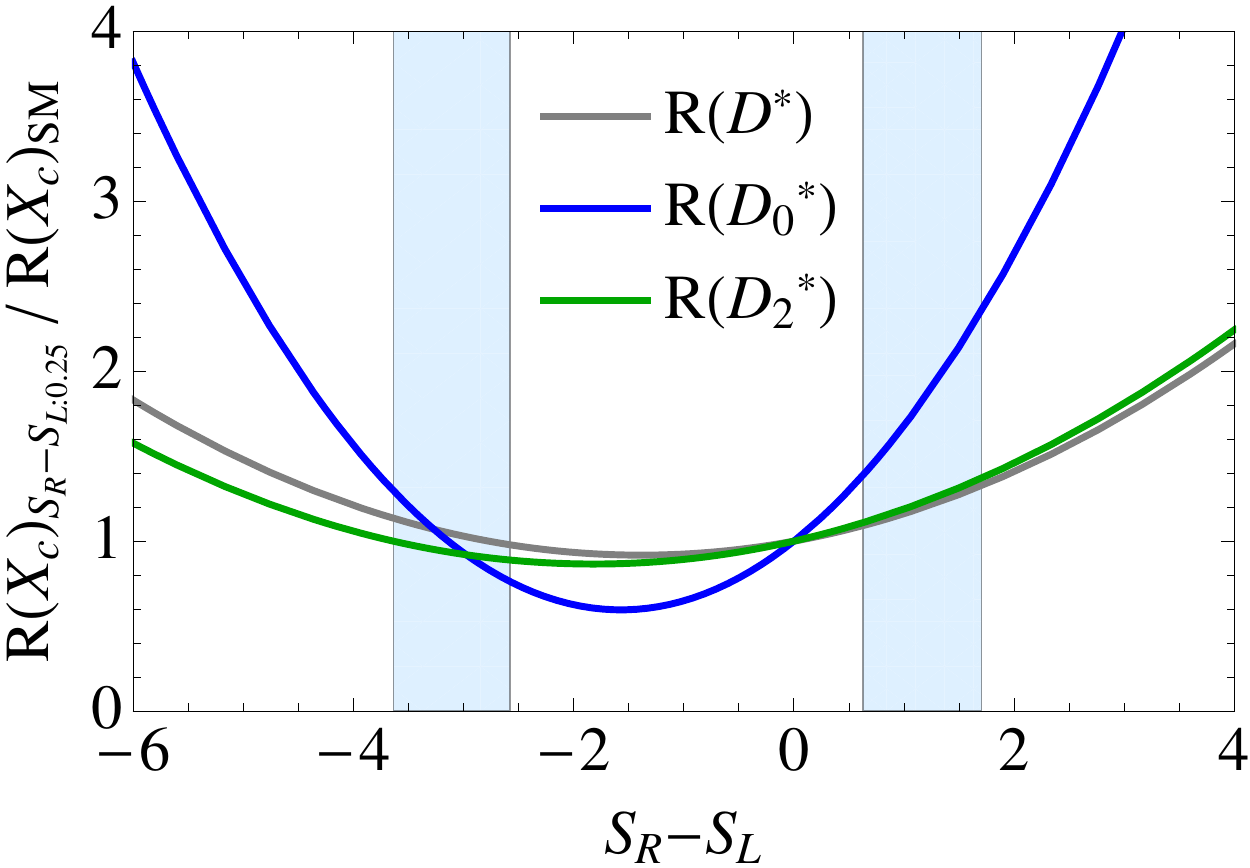}
\caption{Upper plot: Ratios of $\tau$ to light-lepton rates in the type-II 2HDM,
as functions of $\tan\beta/m_{H^\pm}$.  For the four $D^{**}$ states and the two
$D^{(*)}$ mesons $R(X)/R(X)\big|_{\rm SM}$ is shown as functions of
$\tan\beta/m_{H^\pm}$. 
Lower plot: Ratios of $\tau$ to light-lepton rates as functions of $S_R - S_L$,
for $S_R + S_L = 0.25$, for $D^*$, $D_0^*$, and $D_2^*$ final states.  The vertical blue
shaded bands show the allowed regions for $S_R - S_L$ as measured in
Ref.~\cite{Lees:2013uzd}.}
\label{fig:RDds}
\end{figure}

The upper plot in Fig.~\ref{fig:RDds} shows the ratios of $\tau$ to light-lepton
rates as functions of $\tan\beta/m_{H^\pm}$ for the four $D^{**}$ states and for
comparison for the $D^{(*)}$ mesons as well. For the quark masses in
Eq.~(\ref{eq:2hdmtypeIII}) the values of $\ov m_b(m_b) = 4.2$\;GeV and $\ov
m_c(m_b) = 1.1$\;GeV were used. The plot shows for each hadronic final state
$R(X) \big/ R(X)\big|_{\rm SM}$ as a function of  $\tan\beta/m_{H^\pm}$.  While
to such scalar currents the sensitivity of the $B\to D\ell\bar\nu$ appears to be
the best, that is not generic for all new physics scenarios. The lower plot in
Fig.~\ref{fig:RDds} shows the ratios of $\tau$ to light-lepton rates as
functions of $S_R - S_L$ for $S_R + S_L = 0.25$ for the $D^*$, $D_0^*$, and
$D_2^*$ final states.  The rates to the other three states we consider only
depend on $S_R + S_L$ [see Eq.~(\ref{eq:2hdmtypeIII})], so those are not
plotted.  This scenario is motivated by being able to fit, besides
$R(D^{(*)})$, the $q^2$ spectrum measured in Ref.~\cite{Lees:2013uzd} as well. 
The vertical blue shaded bands show the best fit regions~\cite{Lees:2013uzd}. 
Measurements of $R(D^{**})$ can help discriminate between the currently
allowed solutions of $S_L$ and $S_R$, and also distinguish more complex
scenarios.

\section{Summary and Conclusions}
\label{sec:summary}

We performed the first model independent study of semileptonic $B\to D^{**} \ell
\bar\nu$ decays based on heavy quark symmetry, including the full dependence on
the charged lepton mass.  This is important, because future measurements of
$R(D^{**})$ give more complementary sensitivity to new physics than $R(D^{(*)})$.  It
is also important to better understand the semileptonic $B\to D^{**}$ decays in
the zero lepton mass channels, which are significant contributions to the
systematic uncertainties for the measurements of $|V_{cb}|$ and $|V_{ub}|$, in
addition to $R(D^{(*)})$.

There are at least two measurements which could be done with existing data, that
would add substantially to our understanding of $D^{**}$ states and the decays
discussed in this paper: (1) The nonleptonic $B\to D^{**}\pi$ rates have only
been measured with small fractions of the \babar\ and Belle data, and are the
sources of tensions.  Redoing these measurements with the full data sets would
be important. (2) In the strange sector, one should revisit the ratio ${\cal
B}(D_{s0}\to D_s^*\gamma) / {\cal B}(D_{s0}\to D_s\pi)$, for which CLEO obtained
a $\sim$\,3 times stronger upper bound than \babar and Belle, and the latter
experiments have much more data not yet analyzed for this ratio.

Our main results for $R(D^{**})$ are Eqs.~(\ref{eq:AppB_RDds_results}) and the
even more conservative Eqs.~(\ref{eq:AppC_RDds_results}).  The precision of
these predictions can be improved in a straightforward manner in the future,
with more precise measurements of the differential decay rates in the $e$ and
$\mu$ modes.  That will allow to better constrain the (relevant combinations of)
subleading Isgur-Wise functions, thereby reducing the uncertainty of
$R(D^{**})$.  Measuring the $e$ and $\mu$ modes should be high priority also for
their potential impacts on reducing the uncertainties in $|V_{cb}|$ and
$|V_{ub}|$ measurements. 

For the semi-tauonic rate to the sum of four states we obtain ${\cal B} (B\to
D^{**} \tau\bar\nu) = (0.14 \pm 0.03)\%$.  This is smaller than the estimate in
Ref.~\cite{Freytsis:2015qca}; nevertheless, it sharpens the tension between the
data on the inclusive and sum over exclusive $b\to c\tau\bar\nu$ mediated rates.

\acknowledgements

We thank Michele Papucci, Dean Robinson, Marcello Rotondo and Sheldon Stone for
helpful conversations. FB thanks Niklas and Nicole Wicki for good conversations
in Zermatt,  where parts of this manuscript were worked out. Special thanks to
Stephan Duell to point out a typo in the helicity amplitude formulae in the
appendix. FB is supported by DFG Emmy-Noether Grant No. BE 6075/1-1.
ZL thanks the hospitality of the Aspen Center for Physics, supported by the NSF
Grant No.~PHY-1066293. 
ZL was supported in part by the Office of Science, Office of High Energy
Physics, of the U.S.\ Department of Energy under contract DE-AC02-05CH11231.

\appendix

\section{LLSW Form Factor expansion}
\label{App:formfactors}

The used mass splittings and quark masses are listed in
Table~\ref{tab:input_summary}.  The ratios, $\varepsilon_{c,b} =
1/(2 m_{c,b})$, and the subleading Isgur-Wise functions $\tau_{1/2/3}$
also enter the form factor expansion. Here $\tau_{1/2}$ and $\tau_{3/2}$ are the
leading Isgur-Wise function of the $s_l^\pi = \frac{1}{2}^+$ and $s_l^\pi =
\frac{3}{2}^+$ states, respectively.  Below, we repeat for completeness the
expansion of the form factors to order $1/m_{c,b}$~\cite{Leibovich:1997tu,
Leibovich:1997em}.

The form factors for $B \to D_0^*\, \ell\, \bar\nu$ are
\begin{eqnarray}\label{FFD0}
g_+ &=& \varepsilon_c \bigg[ 2(w-1)\zeta_1
  - 3\zeta\, {w\bar\Lambda^*-\bar\Lambda\over w+1} \bigg] \nn\\*
&&{} - \varepsilon_b \bigg[ {\bar\Lambda^*(2w+1)-\bar\Lambda(w+2)\over w+1}\,
  \zeta - 2(w-1)\,\zeta_1 \bigg] , \nn\\*
g_- &=& \zeta + \varepsilon_c \Big[ \chi_{\rm ke}+6\chi_1-2(w+1)\chi_2 \Big] 
  + \varepsilon_b\, \chi_b \,. 
\end{eqnarray}
The form factors for $B \to D_1^*\, \ell\, \bar\nu$ are
\begin{widetext}
\begin{eqnarray}\label{FFD1s}
g_A &=& \zeta + \varepsilon_c \bigg[\frac{w\bar\Lambda^*-\bar\Lambda}{w+1} \zeta
  + \chi_{\rm ke}-2\chi_1 \bigg] - 
  \varepsilon_b\, \bigg[ {\bar\Lambda^*(2w+1)-\bar\Lambda(w+2)\over w+1}\,
  \zeta - 2(w-1)\,\zeta_1 - \chi_b \bigg] ,\nn\\*
g_{V_1} &=& (w-1) \zeta + \varepsilon_c 
  \Big[(w\bar\Lambda^*-\bar\Lambda)\zeta + (w-1)(\chi_{\rm ke}-2\chi_1) \Big] 
  - \varepsilon_b \Big\{\! \big[\bar\Lambda^*(2w+1)-\bar\Lambda(w+2)\big] \zeta 
  - 2(w^2-1) \zeta_1 - (w-1)\chi_b \Big\} , \nn\\*
g_{V_2} &=& 2\varepsilon_c\, (\zeta_1-\chi_2) \,, \nn\\*
g_{V_3} &=& - \zeta 
  - \varepsilon_c \bigg[ {w\bar\Lambda^*-\bar\Lambda \over w+1}\zeta 
  + 2\zeta_1 + \chi_{\rm ke} - 2\chi_1 +2\chi_2 \bigg]
  + \varepsilon_b \bigg[ {\bar\Lambda^*(2w+1)-\bar\Lambda(w+2)\over w+1}\,
  \zeta - 2(w-1)\,\zeta_1 - \chi_b \bigg] . 
\end{eqnarray}

The form factors for $B \to D_1\, \ell\, \bar\nu$ are
\begin{eqnarray}\label{FFD1}
\sqrt6\, f_A &=& - (w+1)\tau 
  - \varepsilon_b \big\{ (w-1) \big[(\bar\Lambda'+\bar\Lambda)\tau 
  - (2w+1)\tau_1-\tau_2\big] + (w+1)\eta_b \big\} \nn\\*
&&{} - \varepsilon_c \big[ 4(w\bar\Lambda'-\bar\Lambda)\tau - 3(w-1) (\tau_1-\tau_2) 
  + (w+1) (\eta_{\rm ke}-2\eta_1-3\eta_3) \big] \,,\nn\\*
\sqrt6\, f_{V_1} &=&  (1-w^2)\tau 
  - \varepsilon_b (w^2-1) \big[(\bar\Lambda'+\bar\Lambda)\tau 
  - (2w+1)\tau_1-\tau_2 + \eta_b \big] \nn\\*
&&{} - \varepsilon_c \big[ 4(w+1)(w\bar\Lambda'-\bar\Lambda)\tau
  - (w^2-1)(3\tau_1-3\tau_2-\eta_{\rm ke}+2\eta_1+3\eta_3) \big] \,, \nn\\
\sqrt6\, f_{V_2} &=& -3\tau - 3\varepsilon_b \big[(\bar\Lambda'+\bar\Lambda)\tau 
  - (2w+1)\tau_1-\tau_2 + \eta_b\big] \nn\\* 
&&{} - \varepsilon_c \big[ (4w-1)\tau_1+5\tau_2 +3\eta_{\rm ke} +10\eta_1 
  + 4(w-1)\eta_2-5\eta_3 \big] \,, \nn\\*
\sqrt6\, f_{V_3} &=&  (w-2)\tau 
  + \varepsilon_b \big\{ (2+w) \big[(\bar\Lambda'+\bar\Lambda)\tau 
  - (2w+1)\tau_1-\tau_2\big] - (2-w)\eta_b \big\} \nn\\*
&& + \varepsilon_c \big[ 4(w\bar\Lambda'-\bar\Lambda)\tau + 
  (2+w)\tau_1 + (2+3w)\tau_2 \nn\\*
&& \quad +(w-2)\eta_{\rm ke} -2(6+w)\eta_1 -4(w-1)\eta_2 -(3w-2)\eta_3 \big] \,. 
\end{eqnarray}
The form factors for $B \to D_2^*\, \ell\, \bar\nu$ are
\begin{eqnarray}\label{FFD2}
k_V &=& - \tau - \varepsilon_b \big[(\bar\Lambda'+\bar\Lambda)\tau 
  - (2w+1)\tau_1-\tau_2 + \eta_b\big] 
  - \varepsilon_c (\tau_1-\tau_2+\eta_{\rm ke}-2\eta_1+\eta_3) \,, \nn\\*
k_{A_1} &=& - (1+w)\tau - \varepsilon_b \big\{ (w-1)
  \big[(\bar\Lambda'+\bar\Lambda)\tau - (2w+1)\tau_1-\tau_2\big] 
  + (1+w)\eta_b \big\} \nn\\*
&&{} - \varepsilon_c \big[ (w-1)(\tau_1-\tau_2)
  + (w+1)(\eta_{\rm ke}-2\eta_1+\eta_3) \big] \,, \nn\\
k_{A_2} &=& - 2\varepsilon_c (\tau_1+\eta_2) \,, \nn\\*
k_{A_3} &=& \tau + \varepsilon_b \big[(\bar\Lambda'+\bar\Lambda)\tau 
  - (2w+1)\tau_1-\tau_2 + \eta_b\big] 
  - \varepsilon_c (\tau_1+\tau_2-\eta_{\rm ke}+2\eta_1-2\eta_2-\eta_3) \,.
\end{eqnarray}

\begin{figure*}[th]
\centerline{
\includegraphics[width=0.48\textwidth]{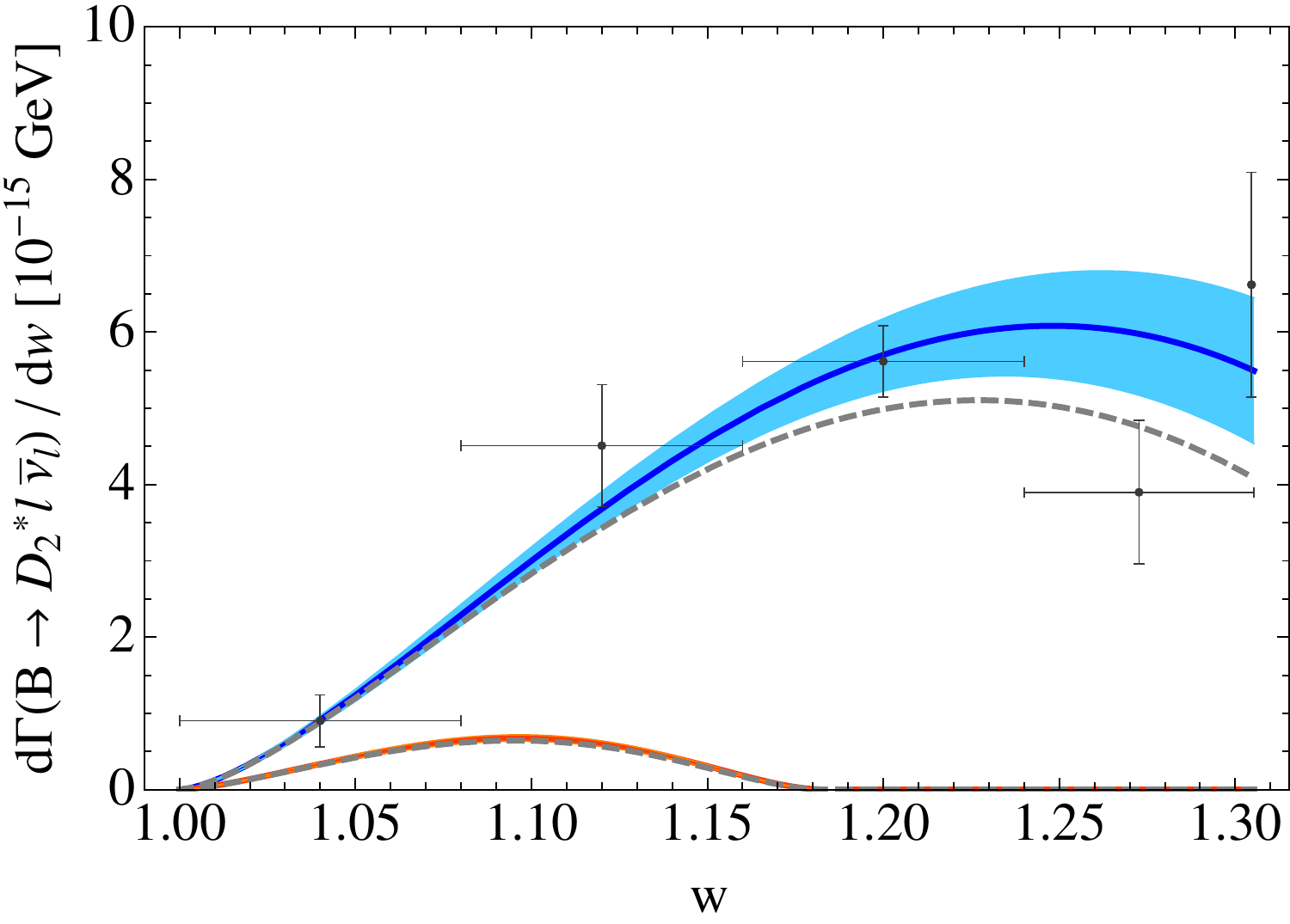}\hfill
\includegraphics[width=0.48\textwidth]{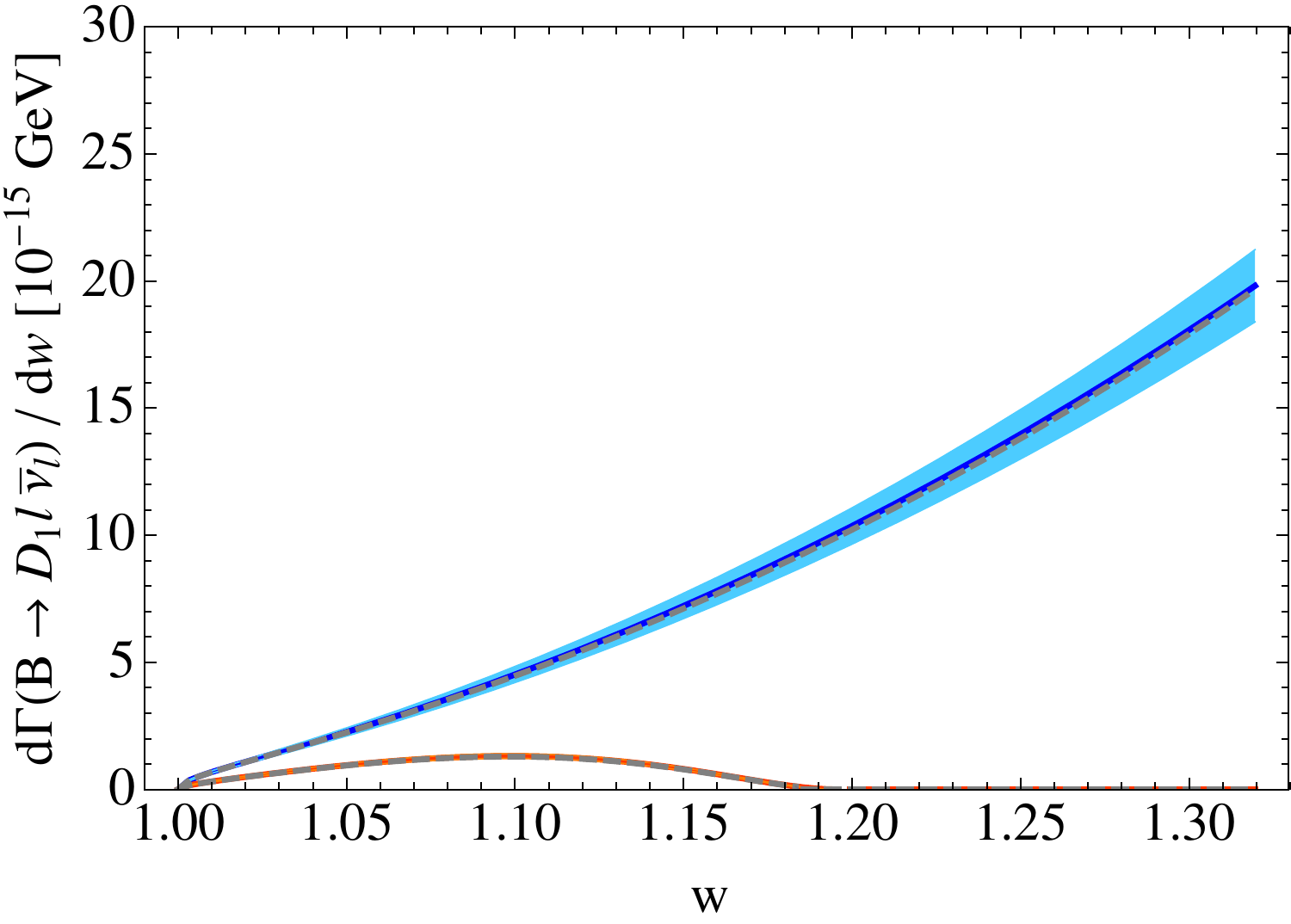} 
}
\centerline{
\includegraphics[width=0.48\textwidth]{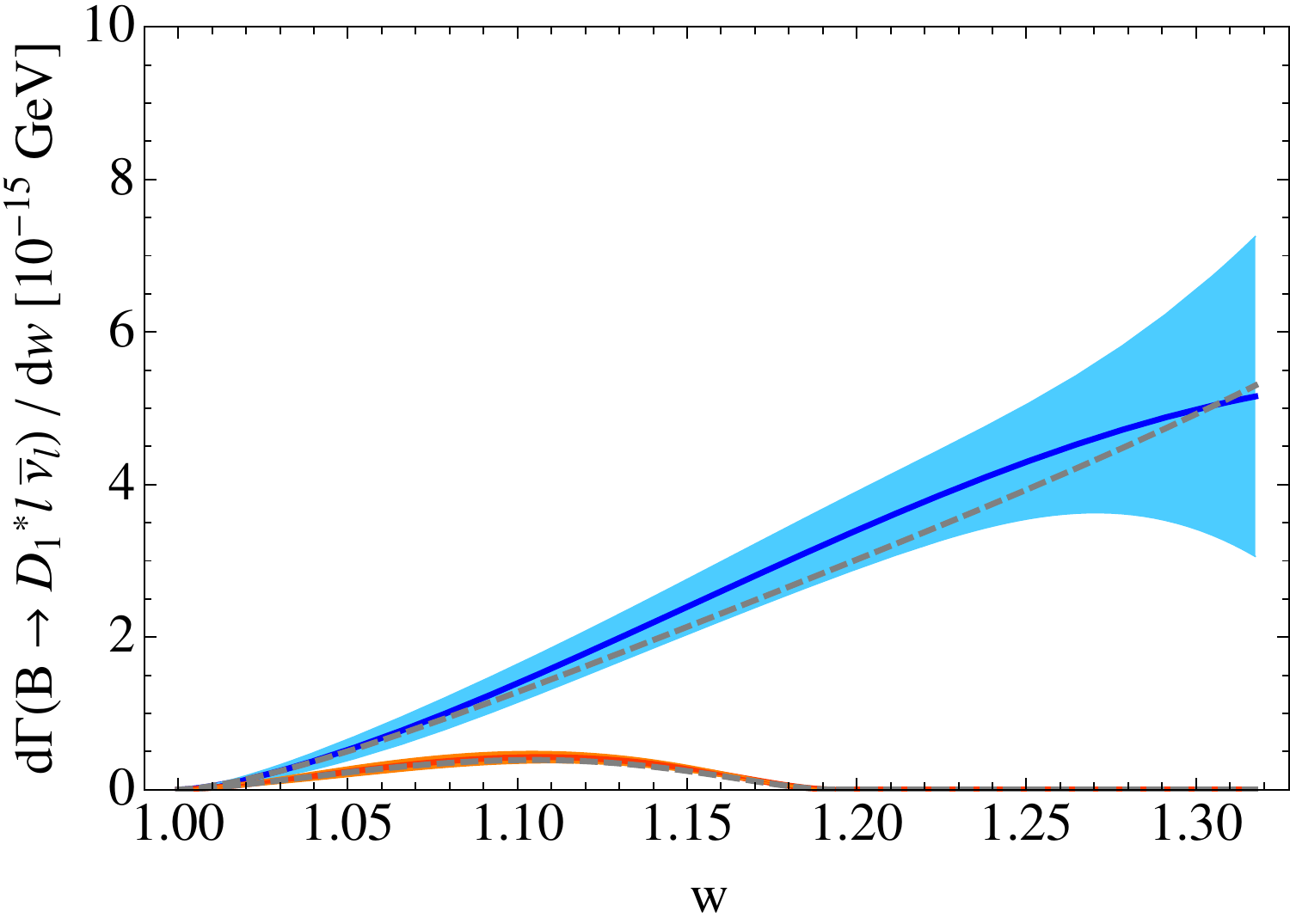}\hfill
\includegraphics[width=0.48\textwidth]{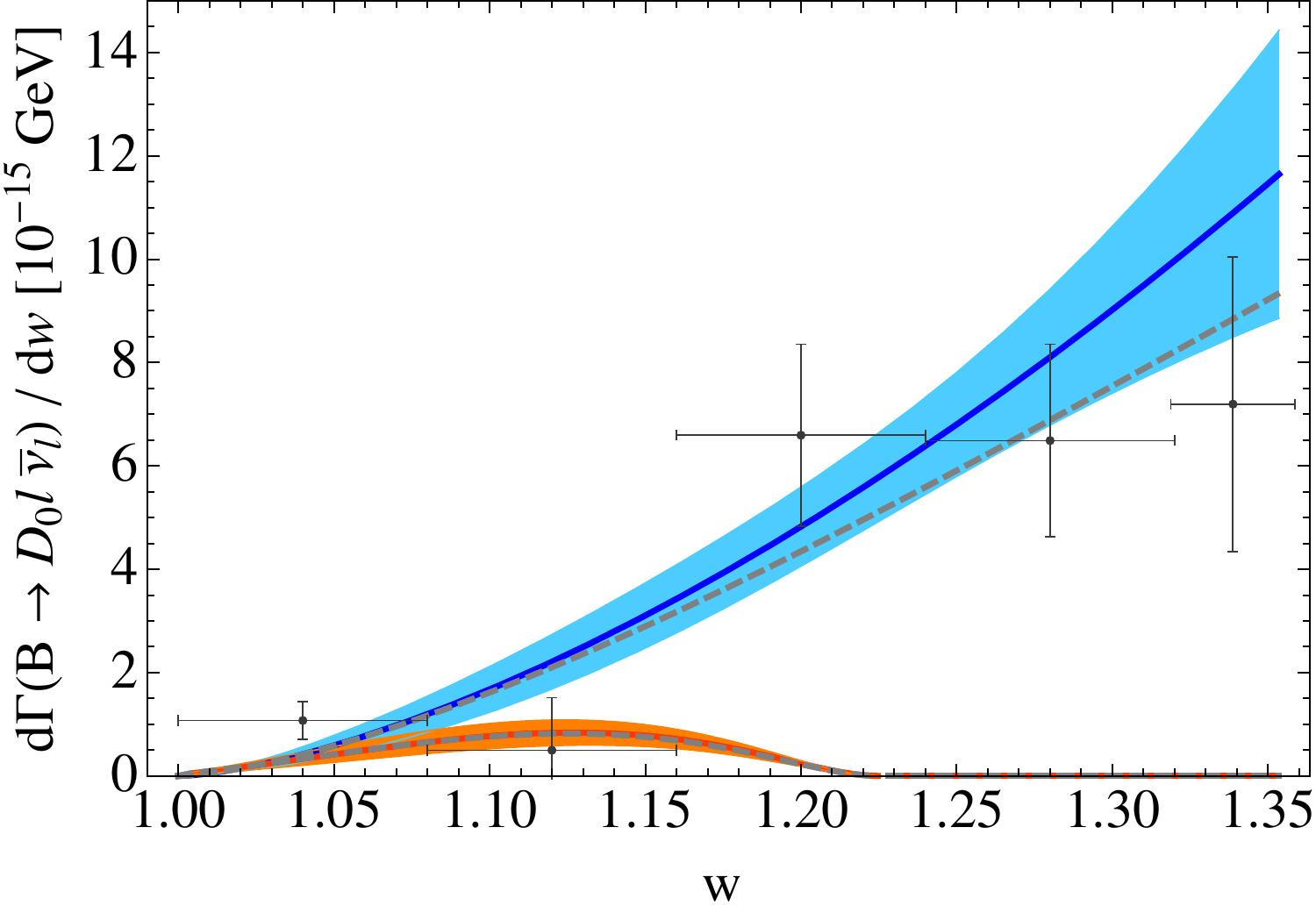}
}
\caption{The colored bands show the allowed 68\% regions for $m_\ell = 0$ (blue)
and $m_\ell = m_\tau$ (orange) for the differential decay rates in \ApproxA. The
dashed curves show the predictions of Ref.~\cite{Leibovich:1997em}. The data
points correspond to the differential semileptonic or nonleptonic branching
fraction measurements described in the text.}
\label{fig:AppA_rates}
\end{figure*}

\section{\ApproxA}
\label{sec:ApproxA}

We attempt to keep the definition as similar to Ref.~\cite{Leibovich:1997em} as
possible.  In light of Eqs.~(\ref{D1rate})--(\ref{D0rate}), we factor out $(1-2r
w+r^2-\rl)^2 / (1-2r w+r^2)^2$, which reduces to 1 in the $\rl\to0$ limit. 
Expanding in powers of $w-1$, we write for the $\frac32^+$ states,
\begin{eqnarray}\label{rt_expn1}
{\d\Gamma_{D_1}\over \d w\,\d\!\cos\theta} &=& 
  \Gamma_0\, \tau^2(1)\, r^3\, \sqrt{w^2-1}\, 
  \frac{(1-2r w+r^2-\rl)^2}{(1-2r w+r^2)^2}\, \sum_n\, (w-1)^n\,
  \bigg\{ \sin^2\theta\, s_1^{(n)} \nn\\*
&&{} + (1-2r w+r^2)\, \Big[ (1+\cos^2\theta)\, t_1^{(n)} 
  - 4\cos\theta\, \sqrt{w^2-1}\, u_1^{(n)} \Big] \bigg\} \,,\\*
{\d\Gamma_{D_2^*}\over \d w\,\d\!\cos\theta} &=& 
   \frac32\,\Gamma_0\, \tau^2(1)\, r^3\, (w^2-1)^{3/2}\,
   \frac{(1-2r w+r^2-\rl)^2}{(1-2r w+r^2)^2}\, \sum_n\, (w-1)^n\,
  \bigg\{ \frac43\,\sin^2\theta\, s_2^{(n)} \nn\\*
&&{} + (1-2rw+r^2)\, \Big[ (1+\cos^2\theta)\, t_2^{(n)} 
  - 4\cos\theta\, \sqrt{w^2-1}\, u_2^{(n)} \Big] \bigg\} \,,
\end{eqnarray}
and for the $\frac12^+$ states,
\begin{eqnarray}\label{D0sexp}
{\d\Gamma_{D_0^*}\over \d w\,\d\!\cos\theta} &=& 
  3\Gamma_0\, \zeta^2(1)\, r^3\, \sqrt{w^2-1}\, 
  \frac{(1-2r w+r^2-\rl)^2}{(1-2r w+r^2)^2}\, \sum_n\, (w-1)^n\,
  \bigg\{ \sin^2\theta\, s_0^{(n)} \nn\\*
&&{} + \Big[(1+\cos^2\theta)\, t_0^{(n)}
  - 4\cos\theta\, \sqrt{w^2-1}\, u_0^{(n)} \Big] \bigg\} \,, \\
{\d\Gamma_{D_1^*}\over \d w\,\d\!\cos\theta} &=& 
  3\Gamma_0\, \zeta^2(1)\, r^3\, \sqrt{w^2-1}\, 
  \frac{(1-2r w+r^2-\rl)^2}{(1-2r w+r^2)^2}\, \sum_n\, (w-1)^n\,
  \bigg\{ \sin^2\theta\, s_{1*}^{(n)}\, \nn\\*
&&{} + (1-2rw+r^2)\, \Big[(1+\cos^2\theta)\, t_{1*}^{(n)}
  - 4\cos\theta\, \sqrt{w^2-1}\, u_{1*}^{(n)} \Big] \bigg\} \,.\label{D1sexp}
\end{eqnarray}
\end{widetext}
The structure of the expansion for $D_0^*$ changes for $\rl \neq 0$ compared to
Ref.~\cite{Leibovich:1997em}, where $(w^2-1)^{3/2}$ occurred before the sum
in the analog of Eq.~(\ref{D0sexp}).  One power of $(w^2-1)$ suppression is
eliminated for $\rl \neq 0$.  In Eq.~(\ref{D0sexp}), $s_0^{(0)}$, $t_0^{(n)}$,
and $u_0^{(n)}$ are proportional to $\rl$, and $s_0^{(1,2)}$ correspond to
$s_0^{(0,1)}$ in Ref.~\cite{Leibovich:1997em}.  For the decays to $D_1^*$,
$D_1$, and $D_2^*$, \ApproxA in this paper coincides with
Ref.~\cite{Leibovich:1997em} in the $\rl\to 0$ limit, while for $D_0^*$ there is
this small difference, which is higher order in $w-1$.\footnote{In
Ref.~\cite{Leibovich:1997em} there is a typo in the $s_0^{(0)}$ coefficient in
Eq.~(3.18): the $4\varepsilon_b \hat\chi_b$ term should read $2\varepsilon_b
\hat\chi_b$, since  $\varepsilon_c \hat\chi_{\rm ke}$ and $\varepsilon_b
\hat\chi_b$ must have the same coefficients.  This is corrected in $s_0^{(1)}$
in Eq.~(\ref{stuD0}) below; note the shift of the upper index by 1, as
explained above.}

The subscripts of the coefficients $s,t,u$ denote the spin of the excited $D$
meson, while the superscripts refer to the order in the $w-1$ expansion.  The
$u_i^{(n)}$ terms proportional to $\cos\theta$ only affect the lepton spectrum,
since they vanish when integrated over $\theta$. (We do not expand the factors
of $\sqrt{w^2-1}$ multiplying these $\cos\theta$ terms.)

We obtain for the coefficients in the $D_1$ decay rate
\begin{eqnarray}\label{stuD1}
s_1^{(0)} &=& 16\varepsilon_c^2\, (\bar\Lambda'-\bar\Lambda)^2\, 
  \Big[ 2(1-r)^2 + \rl \Big]
  + {\cal O}(\rl^2 \varepsilon^2,\, \varepsilon^3)\,, \nn\\*
s_1^{(1)} &=& 12\rl + 32 \varepsilon_c (\bar\Lambda'-\bar\Lambda)
   \big( 1-r^2 + \rl\big) \nn\\*
&+& 8\rl \big[\varepsilon_c (3\hat\eta_{\rm ke} +10\hat\eta_1 -5\hat\eta_3)
  + 3\varepsilon_b \hat\eta_b \big] 
  + {\cal O}(\rl^2 \varepsilon,\, \varepsilon^2) , \nn\\*
s_1^{(2)} &=& 8\, (1+r)^2 + 8\rl\, (2+3\hat\tau')
  + {\cal O}(\rl^2,\, \varepsilon)\,, \nn\\
t_1^{(0)} &=& 16\varepsilon_c^2\, (\bar\Lambda'-\bar\Lambda)^2
  \bigg[2 + \frac{\rl}{(1-r)^2}\bigg]
  + {\cal O}(\rl^2 \varepsilon^2,\, \varepsilon^3)\,, \nn\\*
t_1^{(1)} &=& 4 \bigg[1+\frac{2\rl}{(1-r)^2}\bigg] \big[ 1
  + 2 \varepsilon_c (\hat\eta_{\rm ke} - 2\hat\eta_1 - 3\hat\eta_3)
  + 2\varepsilon_b \hat\eta_b \big] \nn\\*
&&{} + 32 \varepsilon_c (\bar\Lambda'-\bar\Lambda)
  \bigg[1 + \frac{\rl(1+r)}{(1-r)^3}\bigg]
  + 32 \varepsilon_c \rl \frac{4\hat\eta_1 + \hat\eta_3}{(1-r)^2} \nn\\*
&&{}  + {\cal O}(\rl^2\varepsilon,\, \varepsilon^2)\,, \nn\\
t_1^{(2)} &=& 8\,(1+\hat\tau')
  + 16\rl \bigg[\frac{1+\hat\tau'}{(1-r)^2} + \frac{3r}{(1-r)^4}\bigg]
  + {\cal O}(\rl^2,\, \varepsilon)\,, \nn\\
u_1^{(0)} &=& 8\varepsilon_c\, (\bar\Lambda'-\bar\Lambda) 
  \bigg[1-\frac{\rl}{(1-r)^2}\bigg]^2 + {\cal O}(\varepsilon^2) \,, \nn\\
u_1^{(1)} &=& 2 - 4\rl\, \frac{1+r}{(1-r)^3}
  + {\cal O}(\rl^2,\, \varepsilon)\,,
\end{eqnarray}
where $\varepsilon^n$ denotes any term of the form $\varepsilon_c^m
\varepsilon_b^{n-m}$, with $n,\, n-m \geq 0$.
For the decay rate into $D_2^*$, the first two terms in the $w-1$ expansion are
\begin{eqnarray}\label{stuD2}
s_2^{(0)} &=& \big[4(1-r)^2  + \rl\big] \big[1
  + 2\varepsilon_c\,(\hat\eta_{\rm ke}-2\hat\eta_1+\hat\eta_3)
  + 2\varepsilon_b\,\hat\eta_b \big] \nn\\*
&&{} + {\cal O}(\rl^2\varepsilon,\, \varepsilon^2)\,,\nn\\
s_2^{(1)} &=& 4\,(1-r)^2 (1+2\hat\tau') + \rl\, \bigg(\frac72 + 2\hat\tau'\bigg)
  + {\cal O}(\rl^2,\, \varepsilon)\,, \nn\\
t_2^{(0)} &=& 4 \big[1 + 2\varepsilon_b\,\hat\eta_b
  + 2\varepsilon_c (\hat\eta_{\rm ke}-2\hat\eta_1+\hat\eta_3) \big]
  \bigg[1 + \frac{2\rl}{3(1-r)^2}\bigg] \nn\\*
&&{} + {\cal O}(\rl^2,\, \varepsilon)\,, \nn\\
t_2^{(1)} &=& 2 (3 + 4\hat\tau') + 4\rl \bigg[ \frac{(1+r)^2}{(1-r)^4} 
  + \frac{4 \hat\tau'}{3(1-r)^2} \bigg]
  + {\cal O}(\rl^2,\, \varepsilon)\,, \nn\\
u_2^{(0)} &=& 2 - \rl\, \frac{4(1+r)}{3(1-r)^3}
  + {\cal O}(\rl^2,\, \varepsilon)\,. 
\end{eqnarray}
Note that aiming at higher accuracy for the $D_2^*$ rate by keeping the
$s_2^{(2)}$ and $t_2^{(2)}$ coefficients, even at leading order, would introduce
a new parameter, $\tau''(1)$, hence reducing the simplicity of this
approximation (besides deviating from the ``power counting").

For the decay rate into $D_0^*$ we get
\begin{eqnarray}\label{stuD0}
s_0^{(0)} &=& \frac{9\,\rl}2\, (\varepsilon_c+\varepsilon_b)^2\,
  (\bar\Lambda^*-\bar\Lambda)^2 
  + {\cal O}(\rl^2\varepsilon^2,\, \rl\varepsilon^3)\,, \nn\\
s_0^{(1)} &=& \big[ 2(1-r)^2 - \rl \big] \big[ 1
  + 2\varepsilon_c (\hat\chi_{\rm ke} + 6\hat\chi_1 - 4\hat\chi_2) 
  + 2\varepsilon_b \hat\chi_b \big] \nn\\*
&&{} + 6(\varepsilon_c+\varepsilon_b) (\bar\Lambda^*-\bar\Lambda)\, (1-r^2) 
   + {\cal O}(\rl^2\varepsilon,\, \varepsilon^2)\,,\nn\\
s_0^{(2)} &=& (1-r)^2\, (1+4\hat\zeta') - 2\,\rl\, \hat\zeta'
  + {\cal O}(\rl^2,\, \varepsilon)\,. \nn\\
t_0^{(0)} &=& \frac{9\,\rl}2\, (\varepsilon_c+\varepsilon_b)^2\,
  (\bar\Lambda^*-\bar\Lambda)^2 
  + {\cal O}(\rl^2\varepsilon^2,\, \rl\varepsilon^3)\,, \nn\\
t_0^{(1)} &=& \rl\, \big[ 1 
  + 2\varepsilon_c (\hat\chi_{\rm ke} + 6\hat\chi_1 - 4\hat\chi_2)
  + 2\varepsilon_b \hat\chi_b \big] \nn\\*
&&{} + 6\rl\, (\varepsilon_c+\varepsilon_b)\, (\bar\Lambda^*-\bar\Lambda)\,
  \frac{1+r}{1-r}
   + {\cal O}(\rl^2\varepsilon,\, \rl\varepsilon^2) , \nn\\
t_0^{(2)} &=& \rl\, \bigg[ 2\hat\zeta' + \frac{(1+r)^2}{(1-r)^2} \bigg]
  + {\cal O}(\rl^2,\, \rl\varepsilon)\,, \nn\\
u_0^{(0)} &=& - \frac{3\,\rl}{2(1-r)^2}\,
  (\varepsilon_c+\varepsilon_b)\, (\bar\Lambda^*-\bar\Lambda)
   + {\cal O}(\rl^2\varepsilon,\, \rl\varepsilon^2)\,, \nn\\
u_0^{(1)} &=& - \frac{\rl\,(1+r)}{2(1-r)^3}\,
  + {\cal O}(\rl^2,\, \rl\varepsilon)\,. 
\end{eqnarray}
For the decay into $D_1^*$ the coefficients are
\begin{eqnarray}\label{stuD1s}
s_{1*}^{(0)} &=& (\varepsilon_c-3\varepsilon_b)^2\,
  (\bar\Lambda^*-\bar\Lambda)^2 \bigg[(1-r)^2 + \frac{\rl}2 \bigg]
  + {\cal O}(\rl^2\varepsilon^2,\, \varepsilon^3), \nn\\
s_{1*}^{(1)} &=& 3\rl - 2(\varepsilon_c-3\varepsilon_b)\, 
  (\bar\Lambda^*-\bar\Lambda) \big[ (1-r^2) - 2\rl \big]\, \nn\\*
&&{} + 2\rl \big[\varepsilon_c (3\hat\chi_{\rm ke} - 6\hat\chi_1 + 4\hat\chi_2)
  + 3 \varepsilon_b \hat\chi_b \big]
  + {\cal O}(\rl^2\varepsilon,\, \varepsilon^2) , \nn\\
s_{1*}^{(2)} &=& (1+r)^2 + \rl\, ( 2 + 6\hat\zeta')
  + {\cal O}(\rl^2,\, \varepsilon)\,,\nn\\
t_{1*}^{(0)} &=& (\varepsilon_c-3\varepsilon_b)^2\, 
  (\bar\Lambda^*-\bar\Lambda)^2 \bigg[1 + \frac{\rl}{2(1-r)^2} \bigg]
  + {\cal O}(\rl^2\varepsilon,\, \varepsilon^2)\,, \nn\\
t_{1*}^{(1)} &=& \bigg[ 2 + \frac{\rl}{(1-r)^2} \bigg]
  \big[ 1 + 2\varepsilon_c (\hat\chi_{\rm ke}-2\hat\chi_1) 
  + 2\varepsilon_b \hat\chi_b \big] \nn\\*
&&{} + 2 (\varepsilon_c-3\varepsilon_b) (\bar\Lambda^*-\bar\Lambda)
  \bigg[ 2 - \frac{\rl(1+r)}{(1-r)^3} \bigg] \nn\\*
&&{} + \varepsilon_c \frac{8\rl\, \hat\chi_2}{(1-r)^2} 
  + {\cal O}(\rl^2\varepsilon,\, \varepsilon^2)\,, \nn\\
t_{1*}^{(2)} &=& 2 + 4\hat\zeta' + \rl \bigg[ \frac{1+4r+r^2}{(1-r)^4} 
  + \frac{2\hat\zeta'}{(1-r)^2} \bigg] + {\cal O}(\rl^2,\, \varepsilon)\,,\nn\\
u_{1*}^{(0)} &=& (\varepsilon_c-3\varepsilon_b) (\bar\Lambda^*-\bar\Lambda) 
  \bigg[1 + \frac{\rl}{2(1-r)^2} \bigg]
  + {\cal O}(\rl^2\varepsilon,\, \varepsilon^2)\,, \nn\\
u_{1*}^{(1)} &=& 1 - \frac{\rl\,(1+r)}{2(1-r)^3}
  + {\cal O}(\rl^2,\, \varepsilon)\,.  
\end{eqnarray}

Figure~\ref{fig:AppA_rates} compares the differential decay rates in \ApproxA
using the fitted values for the narrow and broad Isgur-Wise function
parametrization in Table~\ref{tab:AppA_results} with
Ref.~\cite{Leibovich:1997em}.

\section{Helicity amplitudes}
\label{App:helamps}

\subsection{Helicity amplitudes for $B \to D_0^*\, \ell \bar\nu$}

The $H_\pm$ helicity amplitudes vanish for semileptonic decays to scalar
final state mesons, and only the zero helicity amplitudes, $H_0$ and $H_t$
contribute to the decay rate,
\begin{align}
H_\pm^S & = 0 \,, \\
H_0^S & = - \frac{m_B \sqrt{r}}{\sqrt{q^2}}\, |\vec p\,'| \bigg[ \bigg( 1 + \frac{1}{r} \bigg) f_+ +  \bigg( 1 - \frac{1}{r} \bigg)f_- \bigg] , \\
H_t^S & = - \frac{m_B \sqrt{r}}{\sqrt{q^2}}\, \big( t_+ \, f_+ + t_- \, f_- \big) ,
\end{align}
with $r = m_{D^{**}} / m_B$, $t_\pm = m_B \mp m_{D^{**}} - E' ( 1 \mp 1/r) = m_B
- E' \mp (m_B\, r- E'/r)$, and
$E'$ denotes the energy of the $D^{**}$ meson in the $B$ rest frame.

\subsection{Helicity amplitudes for $B \to D_1\, \ell \bar\nu$ and $D_1^*\, \ell \bar\nu$}

For vector final state mesons all four helicity amplitudes contribute:
\begin{align} \label{eq:vechelamp}
H_\pm^V & = m_B \sqrt{r}\, f_{V_1} \mp \frac{1}{\sqrt{r}}\,  |\vec p\,'|\, f_{A} \, , \\
H_0^V & = \frac{1}{\sqrt{r\, q^2}} \bigg[ m_B(E' - m_B r^2) \, f_{V_1}
  + \frac{|\vec p\,'|^2}{r} \big( rf_{V_2} +  f_{V_3} \big)  \bigg] , \\
H_t^V & = m_B\, \frac{|\vec p\,'|}{\sqrt{r\, q^2}} \bigg[ f_{V_1}
  + \bigg(1 - \frac{E'}{m_B}\bigg)\, f_{V_2}
  + \bigg(\frac{E'}{m_B\,r} - r\bigg) f_{V_3} \bigg] .
\end{align}
The helicity amplitudes for $D_1^*$ can be obtained by the replacements
$f_{V_1,V_2,V_3,A} \to g_{V_1,V_2,V_3,A}$.

\subsection{Helicity amplitudes for $B \to D_2^*\, \ell \bar\nu$}

For tensor final states also all four helicity amplitudes contribute:
\begin{align}
H_\pm^T & = \mp \frac{1}{\sqrt{2\, r}}\, \frac{|\vec p\,'|^2}{m_B r}\, k_V 
  - \frac{1}{\sqrt{2\, r}} \, |\vec p\,'|\, k_{A_1}\,, \\
H_0^T & = \sqrt{\frac23}\, \frac{|\vec p\,'|}{\sqrt{r^3\, q^2}}
  \bigg[( E' - m_B\, r^2 )\, k_{A_1} \nn\\*
& \qquad + \frac{|\vec p\,'|^2}{m_B} \bigg( k_{A_2} +
  \frac{1}{r} \, k_{A_3} \bigg) \bigg] \,, \\
H_t^T & = \sqrt{\frac23}\, \frac{|\vec p\,'|^2}{\sqrt{r^3\, q^2}}
  \bigg[ k_{A_1} + \bigg( 1 - \frac{E'}{m_B}\bigg) k_{A_2} 
  + \bigg(\frac{E'}{m_B\, r} - r \bigg) k_{A_3} \bigg] \,.
\end{align}

\end{document}